%
\input harvmac.tex
\noblackbox
%


\def\unlockat{\catcode`\@=11}
\def\lockat{\catcode`\@=12}

\unlockat

\def\newsec#1{\global\advance\secno by1\message{(\the\secno. #1)}
\global\subsecno=0\global\subsubsecno=0\eqnres@t\noindent
{\bf\the\secno. #1}
\writetoca{{\secsym} {#1}}\par\nobreak\medskip\nobreak}
\global\newcount\subsecno \global\subsecno=0
\def\subsec#1{\global\advance\subsecno
by1\message{(\secsym\the\subsecno. #1)}
\ifnum\lastpenalty>9000\else\bigbreak\fi\global\subsubsecno=0
\noindent{\it\secsym\the\subsecno. #1}
\writetoca{\string\quad {\secsym\the\subsecno.} {#1}}
\par\nobreak\medskip\nobreak}
\global\newcount\subsubsecno \global\subsubsecno=0
\def\subsubsec#1{\global\advance\subsubsecno by1
\message{(\secsym\the\subsecno.\the\subsubsecno. #1)}
\ifnum\lastpenalty>9000\else\bigbreak\fi
\noindent\quad{\secsym\the\subsecno.\the\subsubsecno.}{#1}
\writetoca{\string\qquad{\secsym\the\subsecno.\the\subsubsecno.}{#1}}
\par\nobreak\medskip\nobreak}

\def\subsubseclab#1{\DefWarn#1\xdef
#1{\noexpand\hyperref{}{subsubsection}%
{\secsym\the\subsecno.\the\subsubsecno}%
{\secsym\the\subsecno.\the\subsubsecno}}%
\writedef{#1\leftbracket#1}\wrlabeL{#1=#1}}
\lockat

%
%
%
\def\CC{{\cal C}}
\def\CE{{\cal E}}
\def\CH{{\cal H}}
\def\CJ{{\cal J}}

\def\CM{{\cal M}}
\def\CN{{\cal N}}
\def\CS{{\cal S}}
\def\CP{{\cal P}}

\def\mod{{\rm mod}}

\def\IZ{\relax\ifmmode\mathchoice
{\hbox{\cmss Z\kern-.4em Z}}{\hbox{\cmss Z\kern-.4em Z}}
{\lower.9pt\hbox{\cmsss Z\kern-.4em Z}}
{\lower1.2pt\hbox{\cmsss Z\kern-.4em Z}}\else{\cmss Z\kern-.4em
Z}\fi}
\def\IB{\relax{\rm I\kern-.18em B}}
\def\IC{{\relax\hbox{$\inbar\kern-.3em{\rm C}$}}}
\def\ID{\relax{\rm I\kern-.18em D}}
\def\IE{\relax{\rm I\kern-.18em E}}
\def\IF{\relax{\rm I\kern-.18em F}}
\def\IG{\relax\hbox{$\inbar\kern-.3em{\rm G}$}}
\def\IGa{\relax\hbox{${\rm I}\kern-.18em\Gamma$}}
\def\IH{\relax{\rm I\kern-.18em H}}
\def\II{\relax{\rm I\kern-.18em I}}
\def\IK{\relax{\rm I\kern-.18em K}}
\def\IP{\relax{\rm I\kern-.18em P}}

\def\inbar{\,\vrule height1.5ex width.4pt depth0pt}
\def\p{\partial}

\font\cmss=cmss10 \font\cmsss=cmss10 at 7pt
\def\IR{\relax{\rm I\kern-.18em R}}
\def\lieg{{\underline{\bf g}}}

\def\ch{{\rm ch}}

\def\Tr{\rm Tr}

\def\zb {{\bar{z}}}

\def\CZ{{\cal Z}}

%
%
\lref\antoni{I. Antoniadis, S. Ferrara, E. Gava, K.S. Narain and T.R. Taylor,
``Perturbative Prepotential and Monodromies in N=2 Heterotic Superstring,''
Nucl. Phys. {\bf B447} (1995) 35, hep-th/9504034.}

\lref\antonii{I. Antoniadis and T.R. Taylor,
``String loop corrections to gauge and Yukawa couplings,''
hep-th/9301033.}

\lref\avatar{A. Losev, G. Moore, N. Nekrasov, S. Shatashvili,
``Four-Dimensional Avatars of 2D RCFT,''
hep-th/9509151.}

\lref\BSV{M. Bershadsky, V. Sadov, and
C. Vafa, ``D-Branes and Topological Field
Theories,'' Nucl. Phys. {\bf B463} (1996) 420; hep-th/9511222.}

\lref\bksdx{T. Banks and L. Dixon, ``Constraints on string vacua with
space-time supersymmetry,'' Nucl. Phys. {\bf B307} (1988) 93.}

\lref\borcha{R. E. Borcherds, ``The monster Lie algebra,'' Adv. Math. {\bf 83}
No. 1 (1990).}

\lref\borchi{R. Borcherds,``Monstrous moonshine
and monstrous Lie superalgebras,'' Invent. Math.
{\bf 109}(1992) 405.}

\lref\borchii{R. Borcherds, ``Automorphic forms
on $O_{s+2,2}(R)$ and infinite products,''
Invent. Math. {\bf 120}(1995) 161.}

\lref\borchiii{R. Borcherds,
``Automorphic forms
on $O_{s+2,2}(R)^+$ and generalized Kac-Moody
algebras,'' contribution to the Proceedings of
the 1994 ICM, Zurich.}

\lref\borchiv{R. Borcherds, ``The moduli space
of Enriques surfaces and the fake monster Lie
superalgebra,''  preprint (1994).}

\lref\borchalg{R. Borcherds, ``Generalized Kac-Moody algebras,'' Journal of
Algebra {\bf 115} (1988) 501.}

\lref\clm{G. L. Cardoso, D. L\"{u}st and T. Mohaupt,
``Threshold corrections and symmetry enhancement in string
compactifications,'' Nucl. Phys. {\bf B450} (1995) 115,
hep-th/9412209.}

\lref\dewit{B. de Wit, V. Kaplunovsky, J. Louis and D.  L\"{u}st,
``Perturbative Couplings of Vector Multiplets in $N=2$ Heterotic String
Vacua,'' Nucl. Phys. {\bf B451} (1995) 53, hep-th/9504006.}

\lref\DoKro{S.K.~ Donaldson and P.B.~ Kronheimer,
{\it The Geometry of Four-Manifolds},
Clarendon Press, Oxford, 1990.}

\lref\eguchi{T. Eguchi, H. Ooguri, A. Taormina and S. K. Yang, ``Superconformal
algebras and string compactification on manifolds with $SU(n)$ holonomy,''
Nucl. Phys. {\bf B315} (1989) 193.}

\lref\egtaor{T. Eguchi and A. Taormina, ``Character formulas for the $N=4$
superconformal
algebra,'' Phys. Lett. {\bf 200B} (1988) 315. }

\lref\fklz{S. Ferrara, C. Kounnas, D. L\"{u}st and F. Zwirner,
``Duality-invariant
partition functions and automorphic superpotentials for $(2,2)$ string
compactifications,''  Nucl. Phys.  {\bf B365} (1991) 431. }

\lref\frenk{I. Frenkel, ``Representations of Kac-Moody algebras
and dual resonance models,'' in {\it Applications
of Group Theory in Physics and Mathematical Physics},
Vol. 21, Lectures in Applied Mathematics, M. Flato,
P. Sally, G. Zuckerman, eds. AMS 1985.}

\lref\golatt{P. Goddard and D. Olive, ``Algebras, Lattices,
and Strings,''
in {\it Vertex operators in mathematics and
physics},'' ed. J. Lepowsky et. al. Springer-Verlag, 1985.}

\lref\kontsevich{Indeed, some aspects of the Dbrane
development were anticipated from this very viewpoint
 in M. Kontsevich, ``Homological Algebra of Mirror
Symmetry,'' Proc. of the 1994 International
 Congress of Mathematicians,
p.120, Birkh\"auser, 1995; alg-geom/9411018}

\lref\mayrst{P. Mayr and S. Stieberger, ``Threshold corrections to
gauge couplings in orbifold compactifications,''  Nucl. Phys. {\bf B407} (1993)
725, hep-th/9303017. }

\lref\mayrsti{P. Mayr and S. Steiberger, ``Moduli dependence of one loop gauge
couplings in $(0,2)$ compactifications,'' Phys. Lett. {\bf B355} (1995) 107,
hep-th/9504129.}

\lref\oogv{H. Ooguri and C. Vafa, ``Geometry of $N=2$ strings,''
Nucl. Phys. {\bf B361} (1991) 969.}

\lref\oogvi{H. Ooguri and C. Vafa,  `` $N=2$ heterotic strings, '' Nucl. Phys.
{\bf B367} (1991) 83.}

\lref\ferrarai{ A.\ Ceresole, R.\ D'Auria, S.\ Ferrara and A.\ Van Proeyen,
``On Electromagnetic Duality in Locally Supersymmetric N=2 Yang--Mills
Theory,''
hep-th/9412200.}

\lref\fhsv{S. Ferrara, J. A. Harvey, A. Strominger, C. Vafa ,
``Second-Quantized Mirror Symmetry, '' Phys. Lett. {\bf B361}
(1995) 59;hep-th/9505162.}

\lref\vaftest{C. Vafa, ``A stringy test of the fate of the conifold,''  Nucl.
Phys. {\bf B447} (1995) 252, hep-th/9505053.}

\lref\gebert{R.W. Gebert,
``Introduction to Vertex Algebras, Borcherds Algebras, and the Monster Lie
Algebra,''
Int. J. Mod. Phys. {\bf A8} (1993) 5441, hep-th/9308151.}

\lref\gnw{R.W. Gebert, H. Nicolai and P.C. West,
``Multistring Vertices and Hyperbolic Kac Moody Algebras,''
hep-th/9505106.}

\lref\hm{J. A. Harvey and G. Moore,
``Algebras, BPS states, and strings,''
hep-th/9510182; Nucl. Phys. {\bf B463}(1996)315.}

\lref\jorgenson{J. Jorgenson and A. Todorov,
``A conjectured analog of Dedekind's eta function
for K3 surfaces,'' Yale preprint. }

\lref\kv{S. Kachru and C. Vafa, ``Exact results for $N=2$ compactifications of
heterotic strings, '' Nucl. Phys. {\bf B450} (1995) 69; hep-th/9505105.}

\lref\lercheeg{W. Lerche, ``Elliptic index and superstring effective
actions,'' Nucl. Phys. {\bf B308} (1988) 102.}

\lref\lsw{see e.g. W. Lerche, A. N. Schellekens and N. P. Warner,
`` Lattices and Strings, '' Phys. Rep. {\bf 177}  (1989) 1.}

\lref\lust{Gabriel Lopes Cardoso ,  Gottfried Curio ,  Dieter Lust ,  Thomas
Mohaupt ,  Soo-Jong Rey, ``BPS Spectra and Non--Perturbative Couplings in N=2,4
Supersymmetric String Theories,''  hep-th/9512129;
Gabriel Lopes Cardoso ,  Gottfried Curio ,  Dieter Lust ,  Thomas Mohaupt,
``Instanton Numbers and Exchange Symmetries in $N=2$ Dual String Pairs,''
hep-th/9603108;
Gabriel Lopes Cardoso ,  Gottfried Curio ,  Dieter Lust,
``Perturbative Couplings and Modular Forms in N=2 String Models with a Wilson
Line,''
hep-th/9608154}

\lref\schwarn{A. N. Schellekens and N. P. Warner, Phys. Lett. {\bf 177B}
(1986) 317; Phys. Lett. {\bf 181B} (1986) 339; Nucl. Phys. {\bf B287} (1987)
317.}

\lref\moorei{ G. Moore,
``Finite in All Directions, '' hep-th/9305139; G.  Moore, ``Symmetries and
symmetry-breaking in string theory,'' hep-th/9308052.}

\lref\mooreii{G. Moore, ``Symmetries of the Bosonic String S-Matrix,''
hep-th/9310026; Addendum to: ``Symmetries of the Bosonic String S-Matrix,''
hep-th/9404025.}

\lref\mukai{S. Mukai, ``Symplectic structure of the
moduli of sheaves on an abelian or K3 surface,''
Invent. Math. {\bf 77}(1984) 101; ``On the moduli space
of bundles on K3 surfaces, I'' in
{\it Vector Bundles on Algebraic Varieties}
Tata Inst. of Fund. Research.}

\lref\nikulini{
V. A. Gritsenko, V. V. Nikulin,
``Siegel automorphic form corrections of some Lorentzian Kac--Moody Lie
algebras, ''
alg-geom/9504006.}

\lref\nikulinii{V. V. Nikulin,
``Reflection groups in hyperbolic spaces and the
denominator formula for Lorentzian Kac--Moody Lie algebras,''
alg-geom/9503003.}

\lref\nikuliniii{
V. A. Gritsenko, V. V. Nikulin, ``K3 Surfaces,
Lorentzian Kac-Moody Algebras, and
Mirror Symmetry,'' alg-geom/9510008.}

\lref\witteg{E. Witten, ``Elliptic Genera and Quantum
Field Theory,'' Commun. Math. Phys. {\bf 109}(1987)525;
``The index of the Dirac operator in loop space,''  Proceedings of the
conference
on elliptic curves and modular forms in algebraic topology, Princeton NJ,
1986.}

\lref\swa{N. Seiberg and E. Witten, ``Electric-magnetic duality, monopole
condensation, and
confinement in $N=2$ supersymmetric Yang-Mills theory,'' Nucl. Phys. {\bf B426}
(1994) 19; (E) {\bf B340} (1994) 485, hep-th/9407087. }

\lref\swb{N. Seiberg and E. Witten, `` Monopoles, duality and chiral symmetry
breaking in $N=2$ supersymmetric QCD, '' Nucl. Phys. {\bf B431} (1994) 484,
hep-th/9408099. }

\lref\dabh{A. Dabholkar and J. A. Harvey, ``Nonrenormalization of the
superstring
tension,'' Phys. Rev. Lett. {\bf 63} (1989) 478; A. Dabholkar, G. Gibbons,
J. A. Harvey and F. Ruiz Ruiz, ``Superstrings and solitons,''
Nucl. Phys. {\bf B340} (1990) 33.}

\lref\sschwarz{A. Sen and J. Schwarz, ``Duality symmetries of 4-D heterotic
strings,''
Phys. Lett. {\bf B312} (1993) 105, hep-th/9305185.}

\lref\hullt{C. Hull and P. Townsend, ``Unity of superstring dualities,'' Nucl.
Phys. {\bf B438} (1995) 109; hep-th/9410167.}

\lref\wittdyn{E. Witten, ``String theory dynamics in various dimensions,''
Nucl.
Phys. {\bf B443} (1995) 85; hep-th/9503124.}

\lref\klti{A. Klemm, W. Lerche and S. Theisen, ``Nonperturbative effective
actions of
$N=2$ supersymmetric gauge theories,'' hepth-9505150. }

\lref\nsi{S. Cecotti, P. Fendley, K. Intriligator and C. Vafa, ``A new
supersymmetric
index, '' Nucl. Phys. {\bf B386} (1992) 405, hep-th/9204102;
S. Cecotti and C. Vafa, ``Ising model
and $N=2$ supersymmetric theories, '' Commun.  Math. Phys. {\bf 157} (1993)
139,
hep-th/9209085.}

\lref\walton{M.  A. Walton,  ``Heterotic string on the simplest Calabi-Yau
manifold and
its orbifold limit, '' Phys. Rev. {\bf D37} (1988) 377. }

\lref\dkl{L. Dixon,  V. S. Kaplunovsky and J. Louis, ``Moduli-dependence of
string
loop corrections to gauge coupling constants, ''Nucl. Phys. {\bf B329} (1990)
27. }

\lref\vadim{V. Kaplunovsky, ``One loop threshold effects in
string unification,'' Nucl. Phys. {\bf B307} (1988) 145,  revised
in hep-th/9205070.}

\lref\kaplouis{V. Kaplunovsky and J. Louis, ``On gauge couplings in string
theory,''
Nucl. Phys. {\bf B444} (1995) 191, hep-th/9502077. }

\lref\cv{E. Calabi and E. Vesentini, Ann. Math. {\bf 71} (1960) 472.}

\lref\FMS{D. Friedan, E. Martinec,  and S. Shenker,
``Conformal Invariance, Supersymmetry, and String Theory,''
Nucl.Phys. {\bf B271} (1986) 93.}

\lref\gilmore{R. Gilmore, {\it Lie Groups, Lie Algebras and some of their
Applications, }Wiley-Interscience, New York, 1974.}

\lref\patera{ J. Patera, R. T. Sharp and P. Winternitz, J. Math. Phys. {\bf 17}
(1976) 1972. }

\lref\agn{I. Antoniadis,  E. Gava, K.S. Narain, ``Moduli corrections to
gravitational
couplings from string loops,'' Phys. Lett. {\bf B283} (1992) 209,
hep-th/9203071; `` Moduli corrections to gauge and gravitational couplings in
four-dimensional superstrings,'' Nucl. Phys. {\bf B383} (1992) 109,
hep-th/9204030.}

\lref\agnt{I. Antoniadis,
 E. Gava, K.S. Narain and T.R. Taylor,
``Superstring threshold corrections to
Yukawa couplings,'' Nucl. Phys {\bf B407} (1993) 706;
hep-th/9212045. Note: These versions are
different.}

\lref\kennati{See K. Intriligator and N. Seiberg, ``Lectures on supersymmetric
gauge
theories and electric-magnetic duality,'' hep-th/9509066 for a recent review.}

\lref\fvanp{S. Ferrara and A. Van Proeyen, `` A theorem on $N=2$ special
K\"ahler product manifolds,'' Class. Quantum Grav. {\bf 6}
(1989) L243. }

\lref\abmnv{L.~Alvarez-Gaum\'e, J.~B.~Bost, G. Moore
P. Nelson, and C.~Vafa,
``Bosonization on higher genus Riemann surfaces,''
Commun. Math. Phys. {\bf112} (1987) 503.}

\lref\GrHa{P.~ Griffiths and J.~ Harris, {\it Principles of Algebraic
geometry}, J.Wiley and
 Sons, 1978. }

\lref\yoshii{H. Yoshii, ``On moduli space of $c=0$ topological conformal
field theories,'' Phys. Lett. {\bf B275} (1992) 70.}

\lref\nojiri{S. Nojiri, ``$N=2$ superconformal topological field theory,''
 Phys. Lett. {\bf 264B}(1991)57. }

\lref\berkvaf{N. Berkovits and C. Vafa, ``$N=4$
topological strings, ''  Nucl. Phys. {\bf B433} (1995) 123, hep-th/9407190.}

\lref\vwdual{ C. Vafa and E. Witten, ``Dual string pairs with $N=1$ and $N=2$
supersymmetry in four dimensions, '' hep-th/9507050. }

\lref\monstref{B. H. Lian and S. T. Yau, ``Arithmetic properties of mirror map
and quantum coupling, '' hep-th/9411234.}

\lref\lianyau{B. H. Lian and S. T. Yau, ``Arithmetic properties of mirror map
and quantum
coupling, '' hep-th/9411234;
``Mirror Maps, Modular Relations
and Hypergeometric Series I ,'' hep-th/9506210;
``Mirror Maps, Modular
Relations and Hypergeometric Series II,''
hep-th/9507153.}

\lref\bcov{M. Bershadsky, S. Cecotti, H. Ooguri and C. Vafa, `` Kodaira-Spencer
theory
of gravity and exact results for quantum string amplitudes, '' Commun. Math.
Phys.
{\bf 165} (1994) 311, hep-th/9309140. }

\lref\douglas{M. Douglas, ``Enhanced Gauge Symmetry in M(atrix) Theory,''
hep-th/9612126}

\lref\fgz{I. Frenkel, H. Garland, and G. Zuckerman, ``Semi-infinite
cohomology and string theory,'' Proc. Nat. Acad. Sci.
{\bf 83}(1986) 8442.}

\lref\gross{ D. Gross, ``High energy symmetries of
string theory,'' Phys. Rev. Lett. {\bf 60B} (1988) 1229.}

\lref\horava{P. Horava, ``Strings on world sheet orbifolds,''  Nucl. Phys.
{\bf B327} (1989) 461; ``Background duality of open string models,'' Phys.
Lett.
{\bf B231}(1989)251.}

\lref\sgntti{M. Bianchi, G. Pradisi and A. Sagnotti, ``Toroidal
compactification and symmetry breaking in open string theories,''
 Nucl. Phys. {\bf B376} (1992) 365.}

\lref\polch{J. Polchinski, ``Combinatorics of boundaries in
string theory,'' Phys. Rev. {\bf D50} (1994) 6041,
hep-th/9407031;  M. B. Green,
``A gas of D instantons,'' Phys. Lett. {\bf B354} (1995) 271,
hep-th/9504108.}

\lref\shenker{S. Shenker, ``Another Length Scale in
String Theory,'' hep-th/9509132.}

\lref\argf{P. C. Argyres and A. E.  Faraggi, ``The vacuum structure and
spectrum of
$N=2$ supersymmetric $SU(n)$ gauge theory, '' Phys. Rev. Lett.
{\bf 74} (1995) 3931, hep-th/9411057.}

\lref\givpor{A. Giveon and M. Porrati, ``Duality invariant string algebra and
$D=4$ effective actions, '' Nucl. Phys. {\bf B355} (1991) 422. }

\lref\dfkz{J. P. Derendinger, S. Ferrara, C. Kounnas and F. Zwirner, ``On loop
corrections to string effective field theories: field-dependent gauge couplings
and
$\sigma$-model anomalies,'' Nucl. Phys. {\bf B372} (1992) 145. }

\lref\agnti{I. Antoniadis, E. Gava, K. S. Narain and T. R. Taylor, ``$N=2$ Type
II-
Heterotic duality and higher derivative F-terms, '' hep-th/9507115.}

\lref\klt{V. Kaplunovsky, J. Louis and S. Theisen, ``Aspects of duality in
$N=2$ string vacua,'' Phys. Lett. {\bf B357} (1995) 71, hep-th/9506110.}

\lref\louispas{J. Louis, PASCOS proceedings, P. Nath ed., World
Scientific 1991.}

\lref\lco{G. L. Cardoso and B. A. Ovrut, ``A Green-Schwarz mechanism
for $D=4$, $N=1$ supergravity anomalies,'' Nucl. Phys. {\bf B369} (1992) 351;
``Coordinate and K\"ahler sigma model anomalies and their cancellation
in string effective field theories,''  Nucl. Phys. {\bf B392} (1993) 315,
hep-th/9205009.}

\lref\levin{L. Levin, Polylogarithms and Associated Functions,
North Holland 1981. See eq.  6.7.}

\lref\witthyper{E. Witten, ``Topological tools in ten dimensional physics, ''
in
{\it Unified String Theories}, eds. M. Green and D. Gross, World Scientific,
Singapore,
1986.}

\lref\fein{A. Feingold and I. Frenkel,
``A Hyperbolic Kac-Moody Algebra and the
Theory of Siegel Modular Forms of Genus
2,'' Math. Ann. {\bf 263} (1083) 87.}

\lref\golatt{P. Goddard and D. Olive, ``Algebras, Lattices,
and Strings,''
in {\it Vertex operators in mathematics and
physics},'' ed. J. Lepowsky et. al. Springer-Verlag, 1985.}

\lref\gebnic{R. W. Gebert and H. Nicolai, `` On $E_{10}$ and the DDF
construction,''
hep-th/9406175.}

\lref\wittorb{E. Witten, ``Space-time and topological orbifolds,'' Phys. Rev.
Lett.
{\bf 61} (1988) 670.}

\lref\fuchs{J. Fuchs, {\it Affine lie algebras and quantum groups,} Cambridge
University Press,  Cambridge, 1992. }

\lref\dkps{M.R. Douglas, D. Kabat, P. Pouliot, and
S.H. Shenker, ``D-branes and short distances in
string theory,'' hep-th/9608024.}

\lref\kac{V. G. Kac, {\it Infinite dimensional Lie algebras,} Cambridge
University Press, Cambridge, 1990. }

\lref\moonshine{J. H. Conway and S. P. Norton, ``Monstrous moonshine,''
Bull. London Math. Soc. {\bf 11} (1979) 308. }

\lref\nikulin{see e.g. V. V. Nikulin, ``Reflection groups in hyperbolic spaces
and the denominator formula for Lorentzian Kac-Moody algebras,
alg-geom/9503003.}

\lref\kmw{V. G. Kac, R. V. Moody, and M. Wakimoto, ``On $E_{10}$,'' In K.
Bleuler
and M. wener, eds. {\it Differential geometrical methods in theoretical
physics.}
Proceedings, NATO advanced research workshop, 16th international conference,
Como
Kluwer, 1988.}

\lref\gebnici{R. W. Gebert and H. Nicolai, ``$E_{10}$ for beginners,''
hep-th/9411188.}

\lref\flm{I. B. Frenkel, J. Lepowsky, and A. Meurman, {\it Vertex operator
algebras
and the monster,} Pure and Applied Mathematics Volume 134, Academic
Press, San Diego, 1988.}

\lref\lianzuck{B. H. Lian and  G. J. Zuckerman, ``New
Perspectives on the BRST Algebraic Structure of
String Theory,''  Commun.Math.Phys. {\bf 154} (1993) 613;
hep-th/9211072.}

\lref\mool{C. Montonen and D. Olive, ``Magnetic monopoles as
gauge particles? ''Phys. Lett. {\bf 72B} (1977)
117; P. Goddard, J. Nuyts and D. Olive, ``Gauge theories
and magnetic charge,''  Nucl. Phys. {\bf B125} (1977) 1.}

\lref\WO{E. Witten and D. Olive, ``Supersymmetry algebras that
include topological charges,'' Phys. Lett. {\bf 78B} (1978) 97. }

\lref\senb{A. Sen, ``Dyon-monopole bound states, selfdual harmonic
forms on the multi-monopole moduli space, and $SL(2,Z)$ invariance
in string theory,'' Phys. Lett. {\bf 329} (1994) 217,
hep-th/9402032.}

\lref\osborn{H. Osborn, ``Topological charges for
$N=4$ supersymmetric gauge theories and monopoles of
spin 1,'' Phys. Lett. {\bf 83B} (1979) 321.}

\lref\andycone{A. Strominger, ``Massless black holes and conifolds in
string theory,'' Nucl. Phys. {\bf B451} (1995) 96; hep-th/9504090.}

\lref\coneheads{B. R. Greene, D. R. Morrison and A. Strominger, ``Black hole
condensation and the unification of string vacua,'' Nucl. Phys. {\bf B451}
(1995) 109; hep-th/9504145.}

\lref\joebrane{J. Polchinski, ``Dirichlet-Branes and Ramond-Ramond charges''
hep-th/9510017.}

\lref\sena{For reviews and further references
see M. Duff, R. Khuri and J. X. Lu,
``String Solitons'' Phys. Rep. {\bf 259} (1995) 213, hep-th/9412184;
 A. Sen, ``Strong-weak coupling duality in four-dimensional string
theory,'' Int. J. Mod. Phys. {\bf A9} (1994) 3707.}

\lref\givrev{A. Giveon, M. Porrati and E. Rabinovici, ``Target space duality in
string
theory,'' Phys. Rep. {\bf 244} (1994) 77;hep-th/9401139.}

\lref\kirka{E. Kiritsis and C. Kounnas, ``Infrared Regularization of
superstring
theory and the one-loop calculation of coupling constants,'' Nucl. Phys.
{\bf B442} (1995) 442, hep-th/9501020.}

\lref\min{J. Minahan, Nucl. Phys. {\bf B298} (1988) 36.}

\lref\afgntrev{I. Antoniadis, S. Ferrara, E. Gava, K. S. Narain and T. R.
Taylor,
``Duality symmetries in $N=2$ heterotic superstring,'' hep-th/9510079.}

\lref\bjulia{B. Julia, in {\it Applications of group theory in physics
and mathematical physics,} ed. P. Sally et. al. (American Mathematical
Society, Providence, 1985). }

\lref\dwvp{B. de Wit and A. Van Proeyen, ``Broken
sigma-model isometries in very special
geometry,'' Phys. Lett. {\bf 293B}(1992)94.}

\lref\klm{A. Klemm, W. Lerche and P. Mayr, ``K3-fibrations and
Heterotic-Type II string duality, '' Phys. Lett. {\bf B357}
(1995) 313, hep-th/9506122.}

\lref\cogp{P. Candelas, X. de la Ossa, P. S. Green and L. Parkes,
``A pair of Calabi-Yau manifolds as an exactly soluble
superconformal theory,'' Nucl. Phys. {\bf B359} (1991) 21.}

\lref\gvz{M. T. Grisaru, A. van de Ven and D. Zanon, Phys. Lett.
{\bf 173B} (1986) 423, Nucl. Phys. {\bf B277} (1986) 388,
Nucl. Phys. {\bf B277} (1986) 409.}

\lref\nakajima{H. Nakajima, ``Homology of moduli
spaces of instantons on ALE Spaces. I'' J. Diff. Geom.
{\bf 40}(1990) 105; ``Instantons on ALE spaces,
quiver varieties, and Kac-Moody algebras,'' Duke Math.
{\bf 76} (1994)365;
``Gauge theory on resolutions of simple singularities
and simple Lie algebras,'' Intl. Math. Res. Not.
{\bf 2}(1994) 61;
``Quiver Varieties and Kac-Moody algebras,''
preprint; ``Heisenberg algebra and Hilbert
schemes of points on projective surfaces,''
alg-geom/9507012; ``Instantons and
affine Lie algebras,'' alg-geom/9502013.}

\lref\nakatsu{ Kazuyuki Furuuchi ,  Hiroshi Kunitomo ,  Toshio Nakatsu,
``Topological Field Theory and Second-Quantized Five-Branes,''
hep-th/9610016}

\lref\gkv{V. Ginzburg, M. Kapranov, and E. Vasserot,
``Langlands reciprocity for algebraic surfaces,''
q-alg/9502013.}

\lref\witzw{E. Witten and B. Zwiebach, ``Algebraic
structures and differential geometry in 2d string theory,''
Nucl. Phys. {\bf B377}(1992)55;hep-th/9201056.}

\lref\cdfkm{P. Candelas, X. de la Ossa, A. Font, S. Katz,
and D. Morrison, ``Mirror Symmetry for Two Parameter
Models - I '' Nucl. Phys. {\bf B416} (1994) 481, hep-th/9308083.}

\lref\iban{G. Aldazabal, A. Font, L. E. Ib\'{a}\~{n}ez and F. Quevedo,
``Chains of $N=2$,$D=4$ heterotic/type II duals,'' hep-th/9510093.}

\lref\wip{Work in progress.}

\lref\martinec{E. Martinec, ``Criticality, Catastrophes, and
Compactifications,'' in L. Brink et. al. eds. {\it Physics and
Mathematics of Strings}, V. Knizhnik memorial volume.
World Scientific, 1990.}

\lref\aspmorr{P.S. Aspinwall and D.R. Morrison,
``String Theory on K3 Surfaces,''  hep-th/9404151.}

\lref\nkawai{T. Kawai,
 ``$N=2$ heterotic string threshold correction,
$K3$ surface and generalized Kac-Moody superalgebra,'' hep-th/9512046; ``String
Duality and Modular Forms,'' hep-th/9607078.}

\lref\henning{
M. Henningson and G. Moore,
`` Counting Curves with Modular Forms,'' hep-th/9602154;
``Threshold corrections in $K3\times T2$
heterotic string compactifications.'' hep-th/9608145. }

\lref\DVV{R. Dijkgraaf,
E. Verlinde, and H. Verlinde, ``Counting
dyons in N=4 string theory,'' hep-th/9607026.}

\lref\sentrick{A. Sen, ``A note on marginally stable bound states in
Type II string theory,'' hep-th/9510229.}

\lref\gottsche{L. G\"ottsche, Lecture Notes in Mathematics 1572,
Hilbert Schemes of Zero-Dimensional Subschemes of Smooth Varieties,
Springer-Verlag, Berlin 1994.}

\lref\ellingsrud{G. Ellingsrud and S. A. Stromme, ``An intersection number
for the punctual Hilbert scheme of a surface,'' alg-geom/9603015.}

\lref\chjp{S. Chaudhuri, C.Johnson and J. Polchinski, ``Notes on
D-branes,'' hep-th/9602052.}

\lref\mukvb{S. Mukai, ``Moduli of vector bundles on K3
surfaces, and symplectic manifolds,''
Sugaku Expositions, Vol. 1, 139.}

\lref\yauzas{S.-T. Yau and E. Zaslow,
``BPS states, string duality, and nodal
curves on K3,'' hep-th/???}

\lref\wittdon{E. Witten, ``Topological quantum field theory,''
Commun. Math. Phys. {\bf 117} (1988) 353.}

\lref\ds{M. Dine,P. Huet and  N. Seiberg, Nucl. Phys. {\bf B322} (1989) 301.}

\lref\givrev{A. Giveon, M. Porrati and E. Rabinovici, ``Target space
duality in string theory,'' Phys. Rept. {\bf 244} (1994) 77.}

\lref\wittenbs{E. Witten, ``Bound states of string and p-branes,''
Nucl. Phys. {\bf B460} (1996) 335; hep-th/9510135.}

\lref\clny{C.G. Callan, C. Lovelace, C. R. Nappi and S. A. Yost, Nucl. Phys.
{\bf B308} (1988) 221.}

\lref\polcai{J. Polchinski and Y. Cai,
``Consistency of open superstring theories,''
Nucl.Phys. {\bf B296} (1988) 91.}

\lref\bib{M. Douglas, ``Branes within branes,'' hep-th/9512077.}

\lref\librane{M. Li, ``Boundary states of D-branes and Dy-strings,''
Nucl. Phys. {\bf B460} (1996) 351, hep-th/9510161.}

\lref\ibrane{M. Green, J. Harvey, and
G. Moore, ``I-brane inflow and anomalous couplings on D-branes,''
hep-th/9605033.}

\lref\dglsmoor{M. Douglas and G. Moore,
``D-Branes, Quivers, and ALE Instantons,''
hep-th/9603167. }

\lref\DRM{D. R. Morrison, ``The geometry underlying
mirror symmetry,'' alg-geom/9608006.}

\lref\segal{G. Segal, ``Equivariant K-theory and
symmetric products,'' manuscript, and talk at
the Aspen Center for Physics, August, 1996.}

\lref\dgm{J. Distler, B. Greene, and D. Morrison,
``Resolving singularities in $(0,2)$ models,''
hep-th/9605222.}

\lref\gothuy{L. G\"ottsche and D. Huybrechts,
``Hodge numbers of moduli spaces of
stable bundles on K3 surfaces,''
alg-geom/9408001.}

\lref\bruzzo{U. Bruzzo and A. Maciocia,
``Hilbert schemes of points on some K3
surfaces and Gieseker stable bundles,''
alg-geom/9412014.}

\lref\donagi{R. Donagi, L. Ein, R. Lazarsfeld,
``A non-linear deformation of the Hitchin
dynamical system,''
alg-geom/9504017.}

\lref\asplouis{P. Aspinwall and J. Louis,
``On the Ubiquity of K3 fibrations in
in string duality,'' Phys. Lett. {\bf B369} (1996) 233; hep-th/9510234.}

\lref\aspenh{P.S. Aspinwall, ``Enhanced
gauge symmetries and Calabi-Yau threefolds,''
hep-th/9511171.}

\lref\kleb{I. Klebanov and A. Polyakov,
``Interaction of discrete states in two-dimensional string theory,''
Mod. Phys. Lett. {\bf A6} (1991)
3273; hep-th/9109032. }

\lref\grndrng{E. Witten, ``Ground ring of two-dimensional string
theory,'' Nucl. Phys. {\bf B373} (1992) 187; hep-th/9108004.}

\lref\vafa{C. Vafa, ``Instantons on D-branes,'' Nucl. Phys. {\bf B463} (1996)
435; hep-th/9512078.}

\lref\gebnic{R.W. Gebert and H. Nicolai,
``An affine string vertex operator construction
at arbitrary level,'' hep-th/9608014.}

\lref\toe{T. Banks, W. Fischler, S. Shenker,
and L. Susskind, to appear.}

\lref\vmii{Vafa and D. Morrison,
``Compactifications of F theory on Calabi-Yau threefolds 2.''
hep-th/9603161.}

\lref\kmv{A. Klemm, P. Mayr, and
C. Vafa, ``BPS States of Exceptional
Non-critical strings,'' hep-th/9607139.}

\lref\kollar{J. Kollar, ``The structure of algebraic
threefolds: an introduction to Mori's program,''
Bull. Am. Math. Soc. {\bf 17}(1987) 211}

\lref\hmii{J. A. Harvey and G. Moore, ``Gravitational threshold corrections
and the FHSV model,'' to appear.}

\lref\lervw{W. Lerche, C. Vafa and N. P. Warner, ``Chiral rings in $N=2$
superconformal theories,'' Nucl. Phys. {\bf B324} (1989) 427.}

\lref\hstrom{J. A. Harvey and A. Strominger, Nucl. Phys. {\bf B449}
(1995) 535, ERRATUM-ibid {\bf B458} (1996) 456; hep-th/9504047.}

\lref\sentoo{A. Sen, Nucl. Phys. {\bf B450} (1995) 103; hep-th/9504027.}

\lref\vwadiab{Cumrun Vafa ,  Edward Witten,
``Dual String Pairs With N=1 And N=2 Supersymmetry In Four Dimensions,''
hep-th/9507050}

%
%

\Title{\vbox{\baselineskip12pt
\hbox{hep-th/9609017}
\hbox{EFI-96-31}
\hbox{YCTP-P15-96}
}}
{\vbox{\centerline{On the algebras of BPS states
} }}

\centerline{Jeffrey A. Harvey}
\bigskip
\centerline{\sl Enrico Fermi Institute, University of Chicago}
\centerline{\sl 5640 Ellis Avenue, Chicago, IL 60637 }
\centerline{\it harvey@poincare.uchicago.edu}
\bigskip
\centerline{Gregory Moore}
\bigskip
\centerline{\sl Department of Physics, Yale University}
\centerline{\sl New Haven, CT  06511}
\centerline{ \it moore@castalia.physics.yale.edu }

\bigskip
\centerline{\bf Abstract}

We define an algebra
on the space of BPS states in  theories with extended supersymmetry.
We show that the algebra of perturbative BPS
states in toroidal compactification of the heterotic string
is closely related to a generalized Kac-Moody
algebra. We use D-brane theory to compare the
formulation of  RR-charged BPS algebras in type II
compactification  with the requirements of
string/string duality and find that the
RR charged BPS states should be regarded
as cohomology classes on   moduli spaces of
coherent sheaves.
The equivalence of the algebra of BPS states in
heterotic/IIA dual pairs elucidates
certain results and conjectures of
Nakajima and Gritsenko \& Nikulin, on geometrically
defined algebras and furthermore
suggests nontrivial generalizations of
these algebras. In particular, to any
Calabi-Yau 3-fold there are two canonically associated
algebras exchanged by mirror symmetry.

\Date{September 1, 1996}
%

\newsec{Introduction}

String theories and field theories with
extended supersymmetry have a distinguished
set of states  in their Hilbert
space known as BPS states. Thanks to
supersymmetry, one can make exact statements
about these magical states
even in the face of all the complexities, perplexities, and
uncertainties that plague most attempts to understand
nonperturbative
Quantum Field Theory and Quantum String Theory. As a result, they
have played a special role in the study of strong-weak
coupling duality in both field theory and string theory.

In this paper we point out that there is a simple, physical,
and universal property of BPS states: They form an
algebra.  There are four reasons the algebra of
BPS states is interesting:

\item{1.} BPS algebras appear to be
infinite-dimensional gauge algebras,
typically spontaneously broken down to
a finite dimensional unbroken gauge
symmetry.

\item{2.} Comparing BPS algebras in
dual string pairs has important applications
in mathematics.

\item{3.} The BPS algebras appear to control
the threshold corrections in
$d=4,\CN=2$ string compactification.

\item{4.} BPS algebras appear to be
intimately related to black hole physics.
In particular, the
counting of nonperturbative black hole
degeneracies seems to be related to
generalized Kac-Moody algebras.

We will discuss $(1)$ and $(2)$ in this paper.
Item $(3)$ is the subject of several papers
\refs{\hm,
\nkawai, \lust, \henning, \hmii
 }.
Item $(4)$  has been proposed recently in an
  imaginative paper of
Dijkgraaf, Verlinde, and Verlinde \DVV.

The outline of this paper is as follows. Section two contains
the basic definition of the algebra of BPS states. In section
three we use the definition to compute the algebra of perturbative
BPS states in toroidal compactifications of heterotic string
theory and
discuss the relation of this algebra to Generalized
Kac-Moody (GKM) algebras.
\foot{In this paper the term GKM is used for something
slightly different from the object defined by Borcherds. See
note added.}
 In the fourth section we turn to an analysis
of BPS states in Type II string theory, we discuss the formulation
in terms of moduli spaces and argue that sheaves provide the
correct language for a general discussion of BPS states. Section 5
develops the sheaf-theoretic interpretation of BPS states in
some detail for $K3$ and $T^4$ compactifications. In section 6
this is extended to the Calabi-Yau case.
Section 7 contains a conjectural method for the computation of the algebra
of  BPS states in Type II string theory. String duality predicts
isomorphisms between certain algebras of Type II BPS states and
the dual algebra of perturbative heterotic BPS states. In sections
8 and 9 we discuss this isomorphism in a certain limit. The final
section contains brief conclusions and a discussion of open issues.

\newsec{The space of BPS states is always an algebra}

\subsec{Definition}

The definition of the algebra of BPS states
uses very little information and is therefore
quite general. We suppose that

\item{1.} There are absolutely conserved
charges $Q$ and therefore the Hilbert space
of asymptotic particle states is graded
$\CH = \oplus \CH^Q$.

\item{2.} In each superselection sector
there is a Bogomolnyi bound on the energy:
\eqn\bogbnd{
E\geq  \parallel \CZ(Q) \parallel
}
where
$\CZ(Q)$ is a central charge and
$ \parallel \cdot \parallel $ is some norm
function.

Given the two conditions above
 we can define the Hilbert space of
BPS states $\CH_{BPS}$ to be the space of
{\it one-particle } states saturating \bogbnd.
\foot{If the one-particle state is
a bound state at threshold it can be
distinguished from a two-particle state
satisfying \bogbnd\ by the representation
of the supertranslation group.}
An algebra is simply a vector space with a
product and we can define the product
\eqn\bpsalg{
\CR: \CH_{BPS} \otimes \CH_{BPS}
\rightarrow \CH_{BPS}
}
as follows. Take two BPS states $\psi_i$ of charges $Q_i $,
$i=1,2$. Boost them by momenta
$\pm \vec p_*$ to produce a two-body state
 in the center of mass frame such that
 the total energy satisfies the
BPS bound $E=\parallel \CZ(Q_1+Q_2)\parallel$.
By definition
{\it $\CR(\psi_1 \otimes \psi_2)$ is
 the orthogonal projection of
$\Lambda_{\vec p_*}(\psi_1)
\otimes \Lambda_{-\vec p_*}(\psi_2)$
onto
$\CH_{BPS}^{Q_1+Q_2}$,
where $\Lambda_{\vec p}$ is
the Lorentz boost.}
The algebra \bpsalg\ is the central object
of study in this paper.

{\bf Remarks}

\item{1.} The nature of the charges and the Bogomolnyi
bound depends on the context (dimension,
number of supersymmetries, global vs. local
supersymmetry, etc.).  For example, in
$d=4,\CN=2$ theories with unbroken
$U(1)^r$ gauge group $Q= (n_I , m^I)$ refers to
the electric and magnetic charges of
the gauge group. The $(\CN=2)$
supersymmetry algebra
in these superselection sectors has a central
charge given in terms of symplectic periods $(X^I,F_I)$ by:
\eqn\cenchrg{
\CZ(Q)  = n_I X^I + m^I F_I
}
In $\CN=2$ supergravity we have
 $ \parallel \CZ\parallel^2\equiv
e^K \vert   \CZ \vert^2$. On the
other hand, for $N=4,8$  $\CZ(Q)$ is the
maximal eigenvalue of the
matrix  $\CZ_{ij}$  etc.

\item{2.} The value of $\vec p_*^2$ is
fixed in terms of central charges:
\eqn\ppole{
\vec p_*^2 = {1 \over 4}\bigl(
\parallel \CZ_{12}\parallel^2
+ {(\parallel \CZ_1\parallel^2 -
\parallel \CZ_2\parallel^2)^2\over
\parallel \CZ_{12}\parallel^2 }\bigr)
-\half (\parallel \CZ_1\parallel^2 +
\parallel \CZ_2\parallel^2)
}
where $\CZ_{12} = \CZ(Q_1) + \CZ(Q_2)$.
It follows from \ppole\ that
 $\vec p_*^2\leq 0$ with
equality iff $\parallel \CZ_{12}\parallel
 =\parallel \CZ(Q_1) \parallel
+ \parallel\CZ(Q_2)\parallel$.
Therefore to implement the above
definition we must use analytic
continuation in $\vec p$. We will
analytically continue in the magnitude,
leaving the direction $\hat p$ real.
Note that the definition of the algebra
in principle depends on the choice of
direction $\hat  p$.\foot{We thank R. Dijkgraaf and E. Verlinde
for emphasizing this.}

\subsec{S-matrix interpretation}

This definition has a simple $S$-matrix
interpretation: Consider the two-body state of two
boosted BPS states $\psi_{1,2}$ with
quantum numbers (in the center of mass
frame)
$(E_i,\pm \vec p;Q_i) $.
Consider the
scattering process:
\eqn\scatter{
\psi_1 + \psi_2 \rightarrow \CF
}
where $\CF$ is some final state which
is a vector in the superselection sector
$  Q_1+ Q_2$.
The $\CS$-matrix for
\scatter\  has a distinguished pole:
\eqn\spole{
\CS(\psi_1 + \psi_2 \rightarrow \CF) \sim {\langle \CF \vert \CR[\psi_1\otimes
 \psi_2] \rangle
\over
s - \parallel \CZ(Q_1+Q_2)\parallel^2}
}
and the residue of the pole defines the product.

In the case of massless BPS states or BPS states
which are bound states at threshold care
is needed. The algebra should be computed using
a limiting procedure, or using techniques such as
those employed in \sentrick.

\subsec{Relation to topological field theory}

The algebra \bpsalg\ may be viewed as a generalization
of the chiral algebra of chiral primary fields in
massive $d=2, \CN=2$ theories \lervw.
Indeed, given the extended spacetime supersymmetry
we may twist the theory \wittdon.
The space $\CH_{BPS}$ is
the BRST cohomology of a scalar supersymmetry
$Q$.  Interpolating operators
which create such states should have nonsingular
products (modulo $Q$).

In many theories BPS states are thought to be smoothly
connected to extremal black hole states. This
suggests an alternative  viewpoint
on the BPS algebra. One could collide two
extremal black holes of charges $Q_1,Q_2$
and consider the amplitude for the resulting
state to settle down to an extremal black
hole of charge $Q_1 + Q_2$. It would be
extremely interesting to relate this amplitude
to the structure constants of the BPS algebra.

\newsec{Example: Toroidally compactified heterotic
string}

The simplest and most elementary
example of the algebra of BPS
states in string theory is given by heterotic string
compactification on $T^d$. This gives the
algebra of  Dabholkar-Harvey (DH) states.
In this case the algebra may be studied
(in string perturbation theory) using
well-developed vertex operator techniques.

The string tree level BPS  algebra is
closely related to the Gerstenhaber and
BV algebras investigated in \lianzuck.
The pole in the S-matrix simply comes from
the
``dot product'' of suitably boosted BRST
classes:
\eqn\dotp{
\CR(V_1\otimes V_2)(z_2, \bar z_2)
\equiv
\lim_{z_1 \rightarrow z_2}
\Lambda_{\vec p_*}\bigl(V_1(z_1,\bar z_1) \bigr)
\Lambda_{-\vec p_*}\bigl(V_2(z_2,\bar z_2) \bigr)
\quad \mod ~ Q_{\rm BRST}
}
where $\Lambda_{\vec p_*}$ is the action of a Lorentz
boost along some fixed direction with magnitude
specified by \ppole.
The BRST class on the LHS of \dotp\ is a class of
(left,right) ghost number
$(2,2)$ but, for the models under consideration, ghost numbers
1 and 2 may be identified via:
\eqn\idengh{
\p c cV \rightarrow c V
}
for $V$ a matter vertex operator.

We now describe the algebra in more detail. BPS states are those
in the right-moving supersymmetric ground state \dabh\ and thus
the space of BPS states has the form
\eqn\dhspace{
\CH_{BPS} = \CH^{\rm mult}\otimes \pi
}
where $\pi$ is a massless  representation of
real dimension
$8_{\rm boson} \oplus 8_{\rm fermion}$
of
the spacetime supertranslation algebra
with $\vec p =0$.
The leftmoving operators also carry spin.
We will show that those multiplets with
no leftmoving spin in the
uncompactified dimensions, $\CH^{\rm mult}_0$
 can be given the structure of
an infinite dimensional
generalized Kac-Moody algebra
\borchalg\borchi.
The full space of multiplets $\CH^{\rm mult}$
carries an interesting algebraic structure,
described below.
 $\CH^{\rm mult}$ is graded by
vectors in the Narain lattice:
\foot{Signature convention: $\Gamma^{16+d,d}$ has
signature $(-1)^{16+d}, (+1)^d$.}
\eqn\grading{
 \CH^{\rm mult} = \oplus_{(P^L;P^R)\in\Gamma^{16+d, d} }
\CH_{(P^L;P^R)}^{\rm mult}
}
Furthermore, we may choose a basis of bosonic states
which (almost) factorize between left and right as:
\eqn\vip{
V_{I,P,\tilde \zeta}=
V^{\rm left}_{P^L,I }(z) \otimes
\tilde V^{\rm right}_{P^R, \tilde \zeta}(\bar z) \epsilon_P
}
where $\epsilon_P$ is a cocycle factor for the lattice.
We will denote right-moving quantities with a
tilde. So,   $\tilde \zeta^M$ is a polarization vector in 10
dimensional space $M=0,...,9$.
While states of string may be associated to
an arbitrary vector in the Narain
lattice $(P_L;P_R)$,  only those charge sectors
with $P^2 \geq -2 $ can satisfy a Bogomolnyi bound.
We thus have:
\eqn\lvlnum{
\eqalign{
N & = \half (   P_R^2 -  P_L^2 )  + 1 = \half P^2 + 1 \geq 0 \cr
E^2 - \vec p^2 & = P_R^2\cr}
}
where $N$ is the oscillator level of
$V^{\rm left}$. The index $I$ runs from
1 to $p_{24}(N)$ over a basis of oscillator
states.

In the basis \vip\ the
product of states in $\CH^{\rm mult}_0$ takes the form
\eqn\prodst{
\CR \bigl[
V_{P_1,I_1,\tilde \zeta_1}\otimes V_{P_2,I_2,\tilde\zeta_2}
\bigr] =
[V^{left}_{P_1,I_1}, V^{left}_{P_2,I_2}] \otimes
\tilde V^{right}_{\tilde\zeta_{12} }
}
where $\tilde\zeta_{12}$ is a function, given below,  of
$\tilde\zeta_i$ and $P_i$, and the first factor is a
{\it Lie bracket}.

To compute the algebra in this basis we first examine the
right-moving component in the $(-1)$ picture:
\eqn\rightop{
\tilde V^{\rm right}_{P^R,\tilde \zeta}(\bar z) =
\tilde c(\bar z) e^{-\tilde \phi} e^{i \tilde k\cdot \tilde x}
 \tilde \zeta\cdot  \tilde \psi  (\zb)
}
Here
$\tilde \phi$ is the superconformal
ghost,
$\tilde x^M$, $M=0,\dots 9$ labels all right-moving
spacetime coordinates. The right-moving momentum
is $\tilde k = (E, \vec p ; P^R) $ and the
BPS condition requires $\vec p=0$ and that
$\tilde k$  is lightlike:
\foot{Our signature convention for the spacetime metric
is that $\eta_{00}=-1$. The index $0$ refers to
time. }
\eqn\rightmom{
\tilde k^2 = 0
}
We also have: $\tilde k\cdot \tilde \zeta =0$ and
$ \tilde \zeta \sim  \tilde \zeta+ \alpha \tilde k$.
The right-moving part of the
product  is most easily computed
by multiplying operators in the $-1$ and $0$ pictures.
the physical boost used in
the definition in the previous section
is characterized by the requirement that
the total spatial momentum vanish and that
the new right-moving momentum
remains lightlike:
\eqn\rigmomi{
(\tilde k_1 (\vec p_*) + \tilde k_2 (-\vec p_*) )^2 = 0 \
}
Under these circumstances the only
BRST invariant operator on the right
is again a d=10 U(1) SYM multiplet. The multiplicative
structure on the multiplet is exactly given
by the on-shell three-point vertices of the SYM
multiplet. In particular for the three-point
vertex of bosons we have:
\eqn\polarz{
\eqalign{
\tilde \zeta_{12}  &=
\biggl[\bigl\{ \tilde \zeta_1(\vec p_*) \cdot \tilde k_2(-\vec p_*) \bigr\}
\tilde \zeta_2 (-\vec p_*) -
\bigl\{\tilde \zeta_2(-\vec p_*) \cdot \tilde k_1(\vec p_*) \bigr\}
\tilde \zeta_1 (\vec p_*)\cr
& - \bigl\{\tilde \zeta_1(\vec p_*) \cdot \tilde \zeta_2(-\vec p_*) \bigr\}
\tilde k_2(-\vec p_*) \biggr]
\mod (\tilde k_1(\vec p_*) + \tilde k_2(-\vec p_*) )\cr}
}
where $\tilde \zeta(\vec p_*)$ is the boosted
polarization tensor.

Now let us turn to the left-moving operators.
These have the form:
\eqn\leftop{
V^{\rm left}_{P^L,I} =
c e^{i k\cdot x (z)   }   \CP_I(\p^* x(z) )
}
where $k=(E,\vec p; P^L)$ is the left-moving
momentum and the BPS condition states that
$\vec p=0$ and
\eqn\leftmom{
\half k^2 + N   = 1 \qquad .
}
In particular,
  the matter part of the vertex operator
is a {\it dimension one primary}.
In \leftop\  $ \CP_I(\p^* x(z) )$ runs over a basis of
representatives of the BRST cohomology so
$I=1, \dots , p_{24}(N)$.  The operator product
of the ghost factors in   two such boosted
states is $c(z_1) c(z_2) = z_{12}\bigl(c\p c + \cdots\bigr) $
and therefore we must isolate the simple pole in the
OPE of two dimension one primaries. Thus the
left-moving matter part of the product
state is given by:
\eqn\ope{
\oint_{z_2}  dz_1
\Lambda_{\vec p_*}\bigl(  V_{P^L_1,I_1}(z_1 ) \bigr)
\Lambda_{-\vec p_*}\bigl(  V_{P^L_2,I_2}(z_2 ) \bigr)
\qquad \mod ~Vir^+
}

It is well-known that the pole terms in
mutually local dimension one primaries
generate a current algebra
\frenk\golatt. (The matter
CFT is not unitary and therefore there will
be higher poles in the OPE, but these do
not contribute to the BRST cohomology. )
It follows immediately that the product in
\ope\ defines a current algebra for the states
which are bound states at threshold.
For these states we can take $\vec p =0$.

When there is a nontrivial binding energy
then $k_1\cdot k_2$ will not be integer and the
simple operator products of DH vertex operators will
not define a new DH state. As for the right-movers,
the boost solves the problem and
\eqn\bsleft{
\eqalign{
k_1(\vec p_*) \cdot k_2(-\vec p_*) & = - P_1\cdot P_2 \cr
& = N_1 + N_2 - N_{12} -1 \cr}
}
where $N_{12}$ is the oscillator level of the product
BPS state.

It remains to determine the properties of the product on states requiring
a boost. For   states in $\CH^{\rm mult}_0$
which are left-Lorentz scalars
 the product again determines a Lie algebra.
This can be seen as
follows. We
choose four left-moving dimension one primaries
$\Psi_i$ and consider   the correlator
\eqn\frpt{
\langle \Lambda_{ 1}(\Psi_1)
\Lambda_{ 2}(\Psi_2)
\Lambda_{ 3}(\Psi_3)
\Lambda_{ 4}(\Psi_4)
\rangle\quad .
}
The $\Lambda_i$ are Lorentz boosts
determined by the conditions that
$(\tilde k_i + \tilde k_j)^2=0$ for
all pairs $i,j$. In terms of the boosts
used to define the product we have
\eqn\bstpair{
\Lambda_i \Psi_i(z_i, \bar z_i)
\Lambda_j \Psi_j(z_j, \bar z_j)
= \Lambda_{ij}\bigl( \Lambda_{\vec p_{ij,*}} \Psi_i
\Lambda_{\vec p_{ji,*}} \Psi_j
\bigr)
}
where $\Lambda_{ij}$ is an overall
Lorentz transformation and $\Lambda_{\vec p_{ij,*}} $
are the special boosts, for the pair $\Psi_i, \Psi_j$
defined above. We are only interested in
the first
order poles in \frpt. Thus we  multiply
\frpt\ by $dz_1\wedge dz_2\wedge dz_3\wedge dz_4$
and consider the resulting expression as a
DeRham class on $(\IP^1)^4 -BD$ where
$BD$ stands for the big diagonal where
any two points coincide. The idea is
that on such a space terms of the
form $z^n dz$ are exact except for $n=-1$.
In this way we focus on the pole terms.

The three point functions are given by
\eqn\thrpt{
\langle \Lambda_{ 1}(\Psi_I)
\Lambda_{ 2}(\Psi_J)
\Lambda_{ 3}(\Psi_K)
\rangle dz_1 \wedge dz_2\wedge dz_3
= f_{IJK}(\hat p_{12})  {dz_1 \wedge dz_2\wedge dz_3
\over z_{12} z_{23} z_{31} }\quad \mod ~ d(*)
}
where we have emphasized  that, having
chosen a basis of states $\Psi_I$, the
 structure constants depend on direction.
The two-point function defines a
positive form on the algebra.
Comparing expressions for \frpt\ derived from
the singularities in $z_1$ with that derived from
the singularities in $z_2$ gives a Jacobi-like
identity on $f_{IJK}$:
\eqn\jaclike{
f_{12}^{~~I}(\hat p_{12}) f_{34 I}(\hat p_{34})
-
f_{13}^{~~I}(\hat p_{13}) f_{24 I}(\hat p_{24})
+
f_{14}^{~~I}(\hat p_{14}) f_{23 I}(\hat p_{23})
 =0
}
where the directions are related to each
other by the requirement that
$\tilde s = \tilde t = \tilde u=0$. For
vertex operators which are scalars under
the left-moving noncompact Lorentz group,
or for states not requiring a boost
the direction dependence vanishes and \jaclike\ is the Jacobi
identity.  In this case the product  defines a
Lie algebra as a subalgebra of the algebra of BPS states.
The general structure on arbitrary BPS states
is given by \jaclike.

Finally, we must  clarify the sense in which
$\CH^{\rm mult}_0$ is a GKM algebra. We would like to apply
the definition
of a GKM algebra in section 4 of \borchi, replacing
only the $\IZ$-grading by a $II^{d+16,d}$-grading.
The positive and negative grading is defined by the
Narain vectors $\pm P$
supporting BPS and anti-BPS states, respectively.
The subspaces at fixed grading with respect to
$II^{d+16,d}$ are finite dimensional. Unfortunately
if we convert the $II^{d+16,d}$-grading to a
$\IZ$-grading the finite-dimensionality of the
graded subspaces need not
hold any longer. Thus, the algebras we are
discussing are themselves generalizations of
generalized Kac-Moody algebras.
\foot{We are grateful to R. Borcherds for
an illuminating comment on this point.}

{}From the construction it is
clear that  the   Lie algebra
$\CH^{\rm mult}_0 \subset \CH^{\rm mult}$ satisfies two
key properties.
It is invariant under
 $O( d+16,d; \IR)$
rotations, and
at enhanced symmetry
subvarieties the massless subalgebra
is the unbroken gauge group of the
low energy theory.
Thus, the BPS algebra is a kind of
universal algebra for toroidal compactification.
We regard it as a physically sensible version of
the ``duality invariant  gauge algebra''
of \givpor\
and of the ``universal gauge algebra''
of \moorei. Note that it involves only
physical on-shell
\foot{Although we use analytic
continuation in $\vec p$, we always
work within the framework of
BRST cohomology.}
states, with positive definite inner
product, and no compactification of
time. (The role of compactified time has
been supplanted by the BPS conditions.)

Because of the properties of the BPS algebra described
above, and
because Narain moduli are Higgs fields,
the algebra of BPS states (or at least $\CH^{\rm mult}_0$)
should be regarded as a
spontaneously broken gauge algebra.

{\bf Remarks}

\item{1.} The above construction is
closely related to the technique used
in \mooreii.
The  boost $\Lambda_{\vec p}$ is similar
to the choice of certain kinematic invariants
in \mooreii. In particular the result of that
paper can be  applied to deduce that the
``Ward identities'' of the algebra of BPS
states completely fixes the tree level
BPS scattering matrix.
 In general, a strongly broken
gauge symmetry is as  useless as having no symmetry
at all, but something unusual appears to be happening
in the present context. Not only is the S-matrix
fixed, but, in   $d=4, \CN=2$ theories the
quantum corrections appear to be closely related
to the BPS algebra \refs{
\hm, \nkawai,\henning, \hmii}.

\item{2.} We have computed the residue using a
tree level string calculation. The nonrenormalization
theorems of string perturbation theory could possibly
be applied here to guarantee that the algebra is
 unchanged to all orders of perturbation theory.

\item{3.} Similar considerations apply to
$d=4,\CN=2$ compactifications of the
heterotic string.
The space of DH states now has the form:
\eqn\dhspce{
\CH_{BPS} = \CH_{vm} \otimes \pi_{vm}
\oplus \CH_{hm} \otimes \pi_{hm}
}
where $\pi_{vm}, \pi_{hm}$ are the massless
vectormultiplet and
hypermultiplet representations of
the supertranslation algebra. Both have
  real superdimension:
$4_{\rm boson} \oplus 4_{\rm fermion} $. (We include the supergravity multiplet
as a vectormultiplet).  The algebra of states now has a
$\IZ_2$ grading with vectormultiplets even and hypermultiplets odd.

\item{4.} Previous attempts \hm\
at defining the BPS algebra
using vertex operators have used the ``left-right swap''
according to which we associate a left-moving current
\eqn\image{
\CJ (z)\equiv e^{i p_L X(z) - i p_R \tilde X(z) } \CP_I(\p^* X)
}
to the internal part of the BPS vertex operator, and then
use null gauging to remove the right-moving oscillators.
We regard the present formulation as a significant
improvement on the old one.

\item{5.} It would be interesting to understand the system
of simple roots for this algebra. The real simple roots
will be the states associated with the $P^2=-2$  vectors.
The set of reflections in these roots generates the
Weyl group of the BPS algebra, which is thus
a  subgroup of the
T-duality group.  This subgroup should be
viewed as  a
gauge group, as in \refs{\ds,\givrev}.

\item{6.}
 A similar construction also applies to
the perturbative BPS states of toroidally compactified
type II strings.
The multiplication $\CR(\psi_1 \otimes \psi_2)$
can be defined, but we lose the obvious connection
to currents since now vertex operators
(for medium sized representations)   satisfy
 $\Delta(\CP)+\half P_L^2- \half P_R^2=0$,
so the algebra need not be a GKM algebra.  This algebra is
of interest because, by $U$-duality, it also computes the
algebra of RR charged BPS states for type II on a torus.

\newsec{Geometrical Realization of BPS states for
type II on Calabi-Yau manifolds $X_d$ }

The results of \hm\ and the previous section exhibit an interesting
algebraic structure in the interactions of BPS states in toroidal
and $K3$ compactifications of the heterotic string. Given the
duality between the heterotic string on $T^4$ and the IIA string
on $K3$ \refs{\hullt,\wittdyn, \sentoo,\hstrom}
and between the heterotic string on $K3 \times T^2$ and
type II theory on K3-fibered Calabi-Yau manifolds \refs{\kv,\fhsv}
one expects to
find the same algebraic structure of BPS states in the dual formulation.
This and the following sections are devoted to the development of
this idea.

We start with type II string theory
on $X_d \times \IR^{D,1}$, where
$D=9-2d$ and $X_d$ is a Calabi-Yau manifold of
complex dimension $d$.  We will be
interested in BPS particle states obtained by
wrapping D-branes on cycles in $X_d$.

\subsec{The generalized Mukai vector}

The charges of the unbroken $U(1)$ gauge
symmetries are naturally associated with a vector:
\eqn\chgelttce{
Q \in H^*(X_d;\IZ)
}
  where $*$ is even for
IIA and odd for IIB strings.
\foot{We assume for simplicity that there is
no torsion. Otherwise we mod out by the torsion.
We will also find a possibility for
fractional charges below, so $\IZ$ should
be replaced by ${1 \over N} \IZ$
where $N$ depends on the manifold $X_d$.}
 The reason is
that the $U(1)$ gauge fields are obtained by
Kaluza-Klein reduction of RR $(p+1)$-form
fields $C^{(p+1)}$ and for each homology
$p$-cycle $\Sigma\subset X_d$ we may
define a $U(1)$ gauge field:
$A_{\Sigma} \equiv \int_\Sigma C^{(p+1)}$.
The charge lattice should have a basis dual to the
basis of gauge fields   hence \chgelttce.
The physical interpretation of $Q$ depends on dimension.
For $D=5$ the particles can only have electric
charge, then $Q\in H^*(K3;\IZ)\cong \Gamma^{20,4}$
is an electric charge vector. For $D=3$
$Q$ is a vector of electric and magnetic
charges.

In many cases the space of BPS states with a
fixed charge can be defined in terms of the
cohomology of the moduli space of instantons \refs{\wittenbs,
\BSV}.   Let us suppose that there
are $r$ wrapped $2d$-branes on $X_d$.
\foot{Here we are considering the IIA theory.}
The low energy
dynamics on the D-brane is governed by
maximally symmetric SYM theory in
$X_d \times \IR$ .
The Chan-Paton spaces
form a vector bundle $E\rightarrow X_d$
and the ten-dimensional gauge field
$A_M$, $M=0, \dots 9$ on $E$ becomes a
gauge field  $A_\mu$,
$\mu=0, \dots 2d$ and a Higgs field $ \Phi_i$
$i=1, \cdots D$ which are $r\times r$ antihermitian
matrices. In general the Chan-Paton
vector bundle $E$ is a twisted vector bundle.
Indeed, the RR charge vector $Q$ and the
characteristic classes of the bundle are
related by the important formula:
\eqn\mukvctor{
\eqalign{
Q = v(E) & \equiv \ch(E) \sqrt{\hat A(X_d)}\cr
& = \ch(E) \sqrt{Td(X_d)}\in H^{2*}(X_d;{1\over N} \IZ)\cr}
}
The second line follows since
$X_d$ is Calabi-Yau. The integer
$N$ depends on $X_d$ and is one for $d=2$,
is a divisor of $24$ for $d=3$, and so on.
The Chern character should be regarded as
$\ch(E)=\Tr \exp \bigl[{1\over 2 \pi} (F-B)\bigr]$ where
$F$ is the field strength of the gauge field on the
brane and $B$ is the bulk NS-NS anti-symmetric
tensor field \wittenbs.
The expression \mukvctor\ gives the various
brane-charges associated to a gauge field
configuration via
$Q=(r_{2d}, r_{2d-2}, \dots, r_0)\in H^0\oplus H^2\oplus
\cdots \oplus H^{2d}$. We will refer to
the vector $v(E)$ as a ``generalized Mukai vector.''

{\bf Remark}. Various pieces of
this formula appeared in \refs{\polcai,\clny,\librane,\bib}
and the final form was derived in \ibrane\
using an anomaly inflow argument.
The vector \mukvctor\ has also appeared
in the mathematics literature in the work of
Mukai  for the case $d=2$ \mukai. The origin of
the vector in \mukai\
is the Riemann-Roch-Grothendieck formula
and is closely related to the derivation of
\ibrane.

\subsec{Attempt at a precise formulation of
the space of BPS states}

In this section we will attempt to give a
precise formulation of the space of
BPS states associated with $r$ wrapped
D-branes on $X_d$.
  Supersymmetric
Yang-Mills does not have nontrivial
infrared dynamics in $\IR^{2d}\times\IR$
for $d\geq 2$. Thus the classical and quantum
moduli spaces must coincide. The situation
can be different for SYM on $X_d\times \IR$
since $X_d$ is compact. However, due to  the
topological nature of BPS states mentioned in
section 2.3,  the space of BPS states
should be independent of the volume
$V$ of $X_d$. This motivates our first
assumption:

{\it Assumption A}: We may describe the
BPS states by the supersymmetric ground
states of a supersymmetric quantum mechanics
with target space given by the moduli of
supersymmetric field configurations.
Note that we have made an important
change of limits, exchanging the large
volume and low energy limits.

 The moduli space of classical ground states
is the moduli space of solutions to the
``generalized Hitchin system'' defined by the
equations:
 \eqn\genhitch{
\delta \chi = \Gamma^{MN} \epsilon_1 F_{MN}
+ 1\epsilon_2 = 0
}
for some pair of  covariantly constant
spinors  $\epsilon_i$ on $\IR^D \times X_d$.
Here $\chi, F_{MN}$ are the gaugino and
field strength respectively taking values in the Lie algebra
$u(r)$. The second term lives in the $u(1)$
subalgebra.
 In $N=4$ SYM with gauge group $U(1)$
the scalar fields can be viewed as the Nambu-Goldstone
modes related to translations transverse to the D-brane, the non-linearly
realized supersymmetry transformation involving $\epsilon_2$ is the
superpartner of these translations.
\foot{The presence of the second
term  was noticed in conversations
with J. Polchinski and A. Strominger.}

We would now like to simplify the
equations \genhitch. Since the gauge
field $A_M$ is a massless field representing
excitations of open strings with
{\it Dirichlet} boundary conditions
we are motivated to make

{\it Assumption B}: The fields
$(A_\mu, \Phi_i)$
in \genhitch\ are functions only of the
coordinates $x^\mu$  on $X_d$, and
not functions of the coordinates  of $\IR^D$.

We will now argue that, while assumptions
A and B are reasonable, and approximately
correct, in fact they are incompatible
with string/string duality.

Let us now examine more closely the
consequences of assumptions A and B.
We can choose our covariantly
constant spinors to be of the
form  $\epsilon \otimes \eta$
where $\eta$ is a constant spinor on $\IR^D$
and $\epsilon$ is covariantly constant on
$X_d$ in the Calabi-Yau metric.
We can normalize $\epsilon_1$ so that
the K\"ahler form is
\eqn\khlrfrm{
\omega_{MN} = \epsilon_1^\dagger \Gamma_{MN}\epsilon_1
}
Expanding,
we find  three terms which must separately vanish.
In the part depending on $\Gamma_{\mu\nu}$
  $\epsilon_2$ is chosen so that the equations
for the Yang-Mills field are:  $F$ is type $(1,1)$
 and
$\omega^{d-1}\wedge F = \lambda \omega^d$,
where $\omega$ is the K\"ahler form  and
$\lambda$ is a constant.  Explicitly,
$\lambda = \int_X \omega^{d-1} c_1/\int_X \omega^d$.
The resulting ``generalized Hitchin equations'' are
\eqna\ghs
$$
\eqalignno{
F& \in \Omega^{1,1}(X_d) & \ghs a\cr
\omega^{d-1}\wedge F & = \lambda \omega^d & \ghs b\cr
D_\mu \Phi_i & = 0 & \ghs c\cr
[\Phi_i, \Phi_j] & = 0 & \ghs d\cr}$$

The   equations \ghs{a,b}\ are the Hermitian
Yang-Mills equations. The $\Phi_i$ represent
the normal motions of the $2d$-brane.
Moreover, from \ghs{c}\ we see that if $\Phi_i$
is nondiagonal then the vector
bundle on $X_d$ must in general be reducible.

For a fixed Mukai vector $Q=v(E)$ will define
$\CM'(Q)$ to be the moduli space of
solutions of \ghs{a,b,c,d}\ modulo the
$U(r)$ gauge group.
We expect that the moduli space $\CM'(Q)$
will be rather singular. But we expect it to be a
stratified space with smooth strata.

Note from \ghs{d}\ that $\CM'(Q)$ has a natural projection
to a configuration space of points.:
\eqn\proj{
\pi: \CM'(Q)  \rightarrow S^r (\IR^D)
}
given by the eigenvalues of the (simultaneously
diagonalizable) $\Phi_i$:
\eqn\eigvls{
\eqalign{
\pi[(A_\mu, \Phi_i)] & \rightarrow \{a^{(1)}_i, \dots a^{(r)}_i \} \cr
 \Phi_i   & \sim  Diag\{ a^{(1)}_i, \dots a^{(r)}_i \}\cr
}
}
These give the positions of the $r$ wrapped branes,
so \proj\ provides a partial stratification of
$\CM'(Q)$ by
$S^r (\IR^D)= \amalg S_\nu^r(\IR^D)$
where $\nu$ labels partitions of $r$.

Finally, from \ghs{a,b}\ we see that over the ``small diagonal''
$\Delta^{(r)} \subset S^r (\IR^D)$ where all points
coincide we have a
moduli space of instantons. We define:
\eqn\cmv{
\CM(Q)\equiv \pi^{-1}(p)
}
 for $p\in \Delta^{(r)}$ (the fiber does not depend on
$p$).  A crucial point is that the space $\CM(Q)$
includes the {\it reducible connections}.
Thus, $\CM(Q)$ will itself be a stratified singular space
with singular strata corresponding to the loci of
reducible connections.
Roughly speaking,
the reducible connections are the connections for which the
gauge field can be made block diagonal:
\eqn\blckdiag{
A = \pmatrix{A^{(1)} & 0 \cr
0 & A^{(2)}\cr}
}
This will happen when we can split the
Chan-Paton bundle as:
$E_3=E_1 \oplus E_2$.
More technically, the holonomy group of the
connection should have trivial normalizer
(in the adjoint group). On the reducible
locus the moduli space is roughly a
product of smaller moduli spaces

Let us now consider the BPS states. These should
correspond to harmonic $L^2$ forms on
$\CM'(Q)$.  Since we are interested in bound states
the forms should have support on the stratum
over $\Delta^{(r)}$.  Moreover, we do not
want wave functions with support on the
reducible locus:  such states will be products
of wave functions already
accounted for at smaller charges and
 do not correspond to
bound states, but rather to two (or more)-particle
states. For this reason we
  expect that the bound states
will be associated with the cohomology of  the
subspace $\CM^{\rm irred}(Q)\subset
\CM(Q)$ of irreducible connections.

With this motivation we therefore adopt
the preliminary definition:
 \eqn\bpsinst{
\CH^Q_{BPS} ~{\buildrel ? \over =}~
 H^*(\CM^{\rm irred}(Q))
}
at least when $r_{2d}>0$. Since
$\CM^{\rm irred}(Q)$ is noncompact
we should specify carefully the notion
of cohomology.
The moduli space has a natural metric, and the
physically correct notion of cohomology
would appear to be $L^2$ cohomology.
\foot{However, G. Segal suggests that
relative cohomology with respect to
the reducible locus might be more
appropriate.}
In fact, as we will see below,
{\it equation \bpsinst\ is incompatible
with string/string duality}. One way to
fix it is discussed in the next section.

Remarks:

\item{1.}  If we have a wrapped $2d$ brane
with a nontrivial normal bundle, then,
as remarked in \BSV\ the scalars
$\Phi_i$ take values in the normal bundle
and we get a topologically twisted SYM.
  In the case of a 2-brane wrapping a
holomorphic curve $\Sigma$ in a K3 surface
the equations \ghs\ become the Hitchin
equations on $\Sigma$, after discarding
the scalars $\Phi $ corresponding to
noncompact directions \BSV.
This is the reason for the terminology
``generalized Hitchin equations.''
Note that the remaining $\Phi_i$ represent
normal motions of the brane within a compact
space, and hence the Hitchin space must
be compactified.

\item{2.} Two  parallel $D$-branes of
spatial dimensions $p,p'$ appear to break
supersymmetry unless $p=p' \mod 4$ \chjp.
This raises a paradox:
How then can there be
$(0,2,4,\dots)$ bound states?
One resolution of this paradox
was explained  in \chjp.
When binding a $p-2$-brane in a $p$-brane
the $p-2$ brane can decay via the nucleation
of magnetic objects. This decay is allowed by
the ``Chern-Simons couplings''
which lead to the formula \mukvctor.
Another
description of the same phenomenon
can be given in terms of the effective field
theory on the $p-2$ brane
\ref\hspm{We learned this during discussions
with J. Polchinski and A. Strominger.}.
Naively there is a negative vacuum energy
for parallel $p$ and $p-2$ branes. However,
the theory develops an FI term and a
hypermultiplet field condenses in such a way
as to maintain supersymmetry.

\item{3.}The above description of BPS states
 differs markedly from the
picture discussed in \dkps. It is likely that
the  two descriptions are valid in different
regimes and that the present one is only valid
in the limit $g_{\rm string}^{II} \rightarrow 0$.
It is quite important to understand this
point more clearly.

\subsec{Chan-Paton sheaves}

Let us return to the proposal
\bpsinst\ for the space of BPS states.
It is easy to see that this is incompatible with
string duality. Consider, for example,
$X_2=K3$, for which IIA is dual to the
heterotic theory on $T^4$. Consider a
state with four-brane charge one and large
$0$-brane charge. There are many heterotic
states with these charges, but there are no
such $U(1)$ instantons.
\foot{This paradox is very similar
to a problem which was addressed in
\avatar, section 5.3,  and the resolution
is the same: one must generalize
from vector bundles to sheaves.}
There are similar problems for other
RR charges.

These problems can be resolved if
we take a compactification of
$\CM^{\rm irred}(Q)$ provided by
moduli spaces of sheaves. That is:
to maintain string duality we should
replace Chan-Paton bundles by
Chan-Paton sheaves. This procedure
restores string
duality and has other advantages,
described below.

The   generalization from a
  twisted  Chan-Paton
vector bundle $E$  to a Chan-Paton sheaf
is extremely natural in D-brane theory. Intuitively,
a sheaf is very much like a vector
bundle except that the dimension of
the fiber can change discontinuously.
In particular, the fibers can be the
zero-dimensional vector space
everywhere except at a point
(``skyscraper sheaves'') or on a
curve. This picture coincides nicely
with the physical picture of the
Chan-Paton vector spaces associated
to 0-branes and wrapped 2-branes, respectively.
We will not go very deeply into
sheaf theory in this paper. Everything
that one needs to know (almost) can
be found in sections 0.3 and 5.3 of
\GrHa.

The introduction of  Chan-Paton sheaves
also provides a nice compactification of
the moduli space of instantons.
We assume here that  $X_d$
is algebraic and hence, by the Donaldson-Uhlenbeck-Yau
theorem the moduli space of irreducible connections
can be identified with the moduli of stable
 holomorphic vector bundles on $X_d$.
In algebraic geometry the natural compactification
of the moduli of holomorphic bundles is
provided by a certain  moduli space of sheaves.

There is another advantage to the sheaf-viewpoint.
When there are no $2d$-branes wrapping
$X_d$, but there are branes wrapping
submanifolds of $X_d$ we cannot use
SYM theory on $X_d$.
The advantage of the sheaf viewpoint
is that in the  IIA theory, the entire
set of $(0,2,4,\dots)$ bound states can
be discussed within a single framework.

{\bf Remarks}:

\item{1.}  An important open problem
is to derive the sheaf viewpoint from first
principles. We believe this can be
done by studying the string field theory of
the DN sector states in D-brane theory.

\item{2.} The necessity to include sheaves has
already
been remarked in another investigation
into D-brane moduli space \dglsmoor\ where
it was
noted that the full equivalence of
instantons with bound states of branes
requires the generalization of vector bundles
to sheaves. We have recently learned that
the idea that the natural setting for D-branes
is in the category of coherent sheaves has
also been advocated by R. Dijkgraaf, M. Kontsevich
\kontsevich,
D. Morrison \DRM,
and G. Segal \segal.
 Sheaves have also recently played
an important role in $(0,2)$
Calabi-Yau compactifications \dgm.

\newsec{Example: Sheaf-theoretic
interpretation of $(0,2,4)$ bound states on
an algebraic $K3$ or abelian surface}

Suppose the compactification manifold
$X_2$ is an algebraic K3 surface or
abelian variety ($T^4$).
\foot{We will
need the technical assumption that
these  surfaces to be
``polarized'' which means that we
choose an embedding of the surface into
projective space. In particular, we assume
the K\"ahler class $[\omega]$ is the
restriction of the hyperplane class.}
In this section we will advocate that, in order
to have string/string duality,
BPS states should be formulated as
cohomology classes on the moduli
space of  so-called
``coherent simple semistable
sheaves'':
\eqn\splmod{
\CH^Q_{BPS} = H^*(\CM^{\rm spl}(Q))
}
Let us briefly indicate why such
animals
should be relevant to D-brane physics.

First,  ``coherent'' essentially
means that the sheaf fits in an exact sequence
$  \CF \rightarrow
\CG \rightarrow \CE\rightarrow 0$
where $\CF, \CG$ are ``locally-free'' -
that is, sheaves of sections of a holomorphic
vector bundle. We will interpret this later
as the condition that the 0-brane and 2-brane states
can participate in interactions with wrapped
4-brane states.

Second, the adjective ``simple'' means that the
sheaf contains no nontrivial
automorphisms
\foot{ Sheaves always carry a trivial
automorphism obtained by simply
scaling all sections by a constant factor.} and
is the analog of the requirement that
the corresponding gauge field be
irreducible. Indeed, note that the direct
sum of two sheaves carries a nontrivial
automorphism obtained by scaling the sections
of  just one summand. The criterion of
simplicity is required if we are
to count BPS bound states, and not
2-particle states at zero relative momentum.
Unfortunately, there are a few expectional
cases, involving  states with zero 4-brane
charge where simplicity is not the correct
criterion. Nevertheless, as described below,
these states still admit a sheaf-description.
\foot{The fact that the sheaves in these
exceptional cases are not simple was pointed
out to us by D. Morrison. }

 Finally
``semistable'' means that for any
subsheaf $0\rightarrow\CF\rightarrow \CE$
we have an inequality on the ``slopes'' $\mu(\CF),
\mu(\CE)$:
\eqn\semistbl{
\mu(\CF) \equiv
{\int \omega^{d-1} c_1(\CF) \over \ch_0(\CF)}
\leq
{\int \omega^{d-1} c_1(\CE) \over \ch_0(\CE)}
\equiv \mu(\CE)
}
(we assume $\ch_0>0$ for simplicity).
The physical interpretation of this is much less
evident, but it is required for a nice
\foot{e.g., Hausdorff} moduli
space. The condition \semistbl\ will play a
role in the geometrical formulation of
the BPS algebra below and therefore has
some physical sense.

Let us work out the
 RR charge vector for
$X_2$ a K3 surface or
abelian variety ($T^4$). Then,
  $p_1=-48$ or $p_1=0$,
respectively and the
BPS states have electric charge vector
given by:
\eqn\mukvct{
\eqalign{
Q& =v(E) = (rk(E)  , c_1(E),
   \epsilon rk(E) +  ch_2(E) )\cr
&
\in H^0(X;Z)  \oplus H^2(X;Z)
\oplus H^4(X;Z)\cr}
}
where $\epsilon=1,0$ for $K3,T^4$.
Note that with our conventions
$\ch_2(E)<0$ for a bundle admitting
an ASD connection.
 The inner product
on a vector $v=(r,c_1,r-\ell)$
is:
\eqn\ipdct{
v^2 = (r,  c_1, r-\ell)^2 =  c_1^2 - 2r (r-\ell) }
where $c_1^2$ is the inner product on
$H^2(X,Z) \cong \Gamma^{19,3}$ or $H^2(X,Z) \cong \Gamma^{3,3}$ for
$K3$ or $T^4$ respectively.
 For $X_2=K3$ we need only interpret
states for $v^2 \geq -2$.
These decompose into BPS and anti-BPS
states.
BPS states have  $r>  0$ or
$r=0, c_1>0$, or $r=c_1=0, \ell>0$.
A  theorem of
Mukai \mukai\ shows that  the space of
coherent simple semistable sheaves with
Chern classes specified by
$Q$ is smooth and compact and
has dimension:
\eqn\dimform{
\dim_{\IR} \CM^{\rm spl} (Q) = 4( \half Q^2 + 1 )
}
This is consistent with string duality.

We will now describe the sheaf-theoretic
interpretation of the various states on
 a case-by-case basis.

\subsec{ $Q=(r;c_1; L)$, $r>1$.}

When $r>1$ there is an open dense set in
$\CM(Q)$ consisting of the moduli space
of holomorphic vector bundles (equivalently,
the moduli of irreducible instantons).
The moduli space is compactified by adding
semistable sheaves.
In the physics literature it is sometimes
stated that $\CM(Q) $ is the
$N$-fold symmetric product
$S^N X_2$ for $N=\half Q^2 + 1$,
or, more accurately, its hyperkahler
resolution by the ``Hilbert scheme of points''
\eqn\eqlspc{
\CM(Q) ~ {\buildrel ? \over = }  ~
X_2^{[N]}
}
 and this is quoted
as providing evidence for string/string duality.
Equation \eqlspc\  is certainly true at the
level of dimensions, by Mukai's theorem,
and is known to
be true at the level of Hodge numbers
for some cases \refs{\gothuy,\bruzzo}
provided we use $\CM^{\rm spl}(Q)$.
It is false at the level of complex
structures, see \mukvb\
for a counterexample. As more detailed
questions about the nature of BPS
states become addressed the exact
nature of the relation of these spaces
will become more important.

\subsec{$Q=(1;0; 1-\ell)$. }

{}From Mukai's theorem
 $\dim \CM^{\rm spl}(Q) = 4\ell$.
Indeed, for this case it is known
that \mukai:
\eqn\ellzer{
\CM^{\rm spl} (Q) \cong   X^{[\ell]}
}
The isomorphism is
explained in the appendix.
More generally, we should modify the above by twisting by
a nontrivial line bundle on $X$
to get charge vector $(1,c_1, 1-\ell)$.  This is covered by
Mukai's theorem, of course.  Note that
\ellzer\ resolves the glaring discrepancy with
string duality noted in section 4.3.

\subsec{ $Q=(0;\ch_1;\ch_2)$ }

Mukai's theorem includes sheaves with
$r=0$ and support on a curve. These are
fairly strange objects from the point of view
of a Yang-Mills theory on $X_2$.  Note that
the curve must be irreducible because we are
only interested in bound states. This will be
guaranteed by the condition that the sheaf
be simple.

Suppose $n$ D-branes wrap a holomorphically
embedded curve $\iota: \Sigma \rightarrow X_2$.
We then have a rank $n$ Chan-Paton vector
bundle $E \rightarrow \Sigma$. The corresponding
sheaf on $X_2$ is
\eqn\curvsky{
\CE= \iota_{ *}(E)
}

The Chern characters of  $\CE$ is easily computed from
the Riemann-Roch-Grothendieck (RRG) theorem:
 \eqn\RRG{
\ch(\CE) Td (TX_2)
=
\iota_*(\ch(E) Td(T\Sigma) )
}
Expanding this out we obtain the Mukai vector:
\eqn\purtwo{
v(\CE) = (0, n[\Sigma], \deg(E) -n \half   \Sigma\cdot\Sigma )
}
where
$\deg E = \int_\Sigma \ch_1(E)$.
The 2-brane states should be
associated to the cohomology
$H^*(\CM^{\rm spl}(v(\CE)))$. Roughly,
$\CM^{\rm spl}(v(\CE))$ is the moduli
of pairs consisting of a holomorphic
curve in a linear system:
$C\subset \vert n [\Sigma]\vert$ together
with a rank $n$ vector bundle
$E\rightarrow C$ of fixed degree.
\foot{There are some exceptional
cases where \purtwo\ does not
define a {\it simple} sheaf, e.g.,
$v(\CE)=(0,L[E],0)$ where $E$ is
elliptic. In these cases there is
still a sheaf-description, similar to
that described in the next subsection. }

In \BSV\  the space of states associated
with wrapped two-branes was characterized in
terms on the  cohomology of
Hitchin moduli space. We need a
unified description of the BPS states
in order to describe the BPS algebras
  so we prefer the sheaf description.
As noted in \BSV\   a paper
of Donagi et. al. \donagi\
shows that there is
close relation between Mukai's moduli
space and Hitchin's.

\subsec{ $Q=(0;\vec 0; -L)$.}

Describing zerobranes of
charge $L$  turns out to be the
most subtle case. They
must correspond to sheaves of length
$L$ but concentrated at one point.\foot{This means
that the
fiber above the point is an $L$-dimensional
vector space, in accord  with
the relation  of $U(L)$ SYM to a
charge $L$ 0-brane.}
At first sight string/string
duality   suggests
that the proper moduli space is the
small diagonal:
$\Delta^{(L)} \subset S^LX$. However, for
the sheaf interpretation we must modify
this slightly.

Let
\eqn\hilsch{
\pi: X^{[L]} \rightarrow S^L X
}
be the Hilbert scheme of points resolving
the symmetric product.  Bound states
will only form when the 0-branes are
at the same point in spacetime, so
we expect the 0-brane charge $-L<0$  states
to be represented by cohomology
classes in
\eqn\invim{
\pi^{-1} (\Delta^{(L)}) = (X^{[L]})_L\equiv \Xi_L
}
There are very natural homology
cycles in $(X^{[L]})_L$.
 Let $\Sigma$ be a cycle in $X$.
Consider the cycle:
\eqn\cycle{
\tilde \Sigma_L \equiv \{ \CS\in
\Xi_L: Supp(\CS) = P \in \Sigma
\}
}
This allows us to map a homology class
$[\Sigma]\in H_*(X)$ to a
 class $[\tilde \Sigma_L]\in H_*\bigl(\Xi_L)$.
The dual cohomology classes to these homology
classes in $\Xi_L$ are the correct
differential forms to associate with pure zerobranes
of charge $-L$.\foot{The space $\Xi_L$ is
topologically complicated. For example,
the fiber has
 lots of cohomology:
$b_{2i}(\pi^{-1}(L[P])) = p(L,L-i)$ \refs{\gottsche,\ellingsrud}
. Nevertheless, there are
distinguished cohomology classes on $\Xi_L$
associated to multiplying the top degree cycle
of the fiber with the homology of the base.
We thank G. Segal for some very helpful
remarks on this point. }
It should be noted that {\it none} of the
sheaves in \invim\ are simple, since they
all have nontrivial automorphisms, given by
arbitrary $GL(L,\IC)$ rotations of the fiber
above a point.

\subsec{Summary}

In conclusion we have shown that
the space of D-brane states
can be interpreted as a space of
 cohomology classes on
the moduli space of coherent simple
sheaves on $X$ if the 2 or 4 brane
charge is positive. For some exceptional
2-brane configurations and for pure
0-branes,
the states are a set of distinguished
cohomology classes in a moduli
space of sheaves  supported on
a curve and on a point, respectively.

\newsec{Calabi-Yau compactification}

If we compactify the IIA string on a
Calabi-Yau threefold $X_3$ then the
RR charged BPS states will be
$(0,2,4,6)$ bound states.
 The RR charge vector is connected to
the characteristic classes of the sheaves
via:
\eqn\cycv{
Q= (\ch_0, \ch_1, \ch_2-{p_1 \over 48}\ch_0 ,
\ch_3-{p_1 \over 48}\ch_1)
 }
$Q$ is now interpreted as
a vector of electric and magnetic charges.
The shift by $-{p_1 \over 48}$ is a geometric
version of the Witten effect. Indeed,
choosing an electric/magnetic polarization
so that  $H^0\oplus H^2$ is the lattice of
magnetic charges we observe a shift in the
electric vector: $q_e \rightarrow q_e
-{p_1 \over 48} q_m$.   In particular,
although the shift induces fractional D-brane
charges, it does not violate the Dirac
quantization condition.\foot{We thank
 M. Douglas and E. Witten for
a discussion on this point.}
Once again we expect
\eqn\bpsagain{
\CH^Q_{BPS}
= H^*(\CM^{\rm spl}(Q))
}
although much less evidence is available to
test this proposal.

\subsec{Special features of K3 fibrations }

In general very little is known about the
BPS algebras associated to general
Calabi-Yau manifolds. However, thanks
to string/string duality it is possible to make
some nontrivial statements about the
algebra in the case that there is a heterotic
dual.

It is now understood that heterotic/IIA
duality is intimately connected
with K3 fibrations \refs{\klm,\vwadiab, \asplouis}.
Let us consider therefore a Calabi-Yau 3-fold
 \eqn\kiiifib{
X_3 \rightarrow \IP^1
}
We denote the K3 fiber over $z\in \IP^1$
by $K_z$.
In all known examples it has Picard number $\geq 1$
on the heterotic side, so $K_z$ is always algebraic.

There is a   subset of states
which are of special relevance in string/string
duality. These are the BPS states which have
finite mass in the heterotic weak coupling limit:
\eqn\finmass{
Im(t_s) = \int_{\IP^1} \omega \rightarrow \infty
}
These will be $(0,2,4)$ bound states based on supersymmetric
cycles which only wrap in the fiber $K_z$.
We will refer to these as the ``fiber bound states.''

Let us   determine the lattice of charges and
RR charge vectors of the fiber bound states.
Consider first the supersymmetric
2-cycles in the fiber. These
 must be holomorphic curves in
the CY $X_3$. Therefore, the supersymmetric 2-cycles
with finite mass in the limit
\finmass\    must be holomorphic
curves in the $K3$ {\it in the complex structure
of $K_z$}. The elementary
2-branes will be labelled by vectors
$r\in \Delta^{ir} \subset H^2(K_z;\IZ)$
of classes dual to irreducible holomorphic
curves. Note that
  $r^2= 2g-2 \geq -2$ determines the
genus of the curve. Multiply wrapped
branes with 2-brane charge $Nr$
can produce new bound states. Therefore,
the set of 2-brane charges is contained
in the NEF cone $NEF(K_z)$
and correspondingly, the lattice of
2-brane charges is a sublattice of the Picard lattice
$Pic(K_z)$.
(For a clear discussion of these concepts in
the physics literature see
\refs{\asplouis,\aspenh}.)
In fact, the Picard lattice can
undergo monodromy when circling
a singular fiber. Thus, in general,
we expect the lattice of 2-brane
charges to be $Pic(K_z)^{\rm invt}$
\aspenh.

To obtain the full lattice of charges
we recall that there are
  2 more gauge fields from
wrapping $C^{(5)}$ on the K3 surface and
from $C^{(1)}$, giving
another lattice $\Gamma^{1,1}$ for
$H^0 (K_z;\IZ) \oplus H^4 (K_z;\IZ) $. Note
that rather than use the K\"ahler class of
the $\IP^1$ base we have used it's
magnetic dual  corresponding to
wrapping $C^{(5)}$ on the K3 fiber.
Thus we are {\it not} using the
standard CY polarization $H^0\oplus H^2$
for the magnetic charges.
\foot{and since the intersection form on
$H^{2*}(X_3)$ is {\it symmetric} it is a little
mysterious why, on {\it a priori} grounds,
 such distinct polarizations
should be related by electro-magnetic duality.}
The  total lattice of charges is
\eqn\fiberkiii{
  (r_4, r_2,r_0)\in H^0(K_z;\IZ)
\oplus Pic(K_z)^{\rm invt} \oplus H^4(K_z;\IZ)
}
This lattice embeds into the full lattice $H^*(X_3)$.
The moduli space of sheaves $\CM^{\rm spl}(Q)$ should
be regarded as the moduli of sheaves in the full
Calabi-Yau with specified Chern classes.

We may now compare with the predictions of
string/string duality.
On the heterotic side the  lattice
of charges is the lattice of vectormultiplet charges in
a heterotic dual theory.  Here the gauge instanton
breaks $E_8 \times E_8$ to a rank $s$ subgroup
leaving a Narain moduli space based on a lattice
$\Gamma^{s+2,2}$.  By general results $Pic(K_z)$ is a
lattice of signature $[(-1)^n, (+1)^1]$. We identify
 $Pic(K_z)= \Gamma^{s+1,1}\subset \Gamma^{s+2,2}$.
By comparing this to the Type II picture and
using string duality we see that the 2-brane charges
must in fact fill the entire (monodromy
invariant) NEF cone.  Moreover,
the perturbative BPS states
in the heterotic description (which correspond to
\finmass), are easily described in
terms of vertex operators, and are counted by
the elliptic genus of certain vector bundles on
the heterotic $K3$ surface \hm.

Remarks:

\item{1.} It is interesting to contrast
  \fiberkiii\ with the lattice of
2-brane charges in K3 compactification.
In K3 compactification we do not
count holomorphic curves in $K3$,
but, rather, curves which are holomorphic
in {\it some} complex structure
compatible with the fixed hyperk\"ahler
structure. The appropriate
counting function is $1/\eta^{24}$ \BSV.
By contrast, the holomorphic curves in a
fixed family of complex structures is a more
subtle object. For example, in \hm\ the
counting function for the K3 family
$x_1^2 + x_2^3 + x_3^7 + x_4^{42} =0$
was identified as $E_6/\eta^{24}$.

\item{2.}  A related point is that
in the heterotic dual the BPS
spectrum is {\it chaotic} in the sense that
it changes discontinuously on a dense subset of
hypermultiplet moduli space \hm. This is most
easily seen in the heterotic dual where it
corresponds to discontinuity as functions of
the hypermultiplets describing the heterotic
K3 moduli. Translating this
to the type IIA side we see that the
BPS spectrum is highly chaotic as a function
of the complex structure. This makes sense: a
small perturbation of the complex structure
changes wildly the allowed holomorphic curves
in the K3 surface (generically there are none,
even for an algebraic K3 the Picard lattice
jumps discontinuously). In spite of this chaotic
structure, the difference between the number of
vector multiplet and hyper multiplet BPS states
is stable and it is this difference that governs
physical quantities \hm.

\newsec{The geometrical realization of the
BPS algebra for type II strings}

We would like to calculate the bound state BPS pole
in the  scattering of two BPS states to   a final
state $\CF$:
\eqn\scat{
\psi_1 + \psi_2 \rightarrow \psi_3
\rightarrow \CF
}
corresponding to   charge vectors:
\eqn\scati{
Q_1 + Q_2 = Q_3
}

\subsec{Positive and negative BPS states}

It is important to note that we are scattering
both  BPS and anti-BPS states.
The distinction
is determined by the orientation of $X_d$.
We refer to
these as positive and negative BPS states.
The notion of positivity corresponds to
the positivity of roots in a Lie algebra, and
plays an important
role in the geometrical formulation of
the BPS algebra.

{\bf Example 1}: In the case of 0,2,4 bound states on K3
we can order the charges by saying that

1. $(r,\vec c_1, r- \vert ch_2 \vert ) > 0 $ if $ r >0$

2. $(0, \vec c_1, -\vert ch_2\vert ) >0 $ if $\vec c_1 = \sum n_i [\Sigma_i]$
is in the NEF cone, i.e., if $n_i$ are nonnegative.

3. $(0; 0; -L ) >0$ for $L>0$

{\bf Example 2}:
We now consider $(0,2,4,6)$ bound states
on a Calabi-Yau 3-fold $X_3$.
These will have  charges
 $(r_6,r_4,r_2, r_0) \in H^*(X;Z)$.
$r_2, r_4$ are vectors in lattices, but these
lattices still have cones: the Mori cone and
the NEF cone respectively.
\foot{The Mori cone is the
cone in $H_2(X;Z)$ of homology
classes of holomorphic
curves in $X_3$
\kollar.}
Bogomolnyi states still lie in a cone in
these lattices and
we say that the BPS states are positive if

$r_6 >0$ or,

$r_6=0$, $r_4>0$ or,

$r_6=r_4=0$, $r_2>0$, or,

$r_6=r_4=0=r_2=0$, $r_0>0$

\subsec{The correspondence conjecture}

We now consider the BPS states as differential
forms on the moduli space of sheaves.
Then the BPS states are represented by
cohomology classes
$$
\omega_i \in H^*(\CM^{\rm spl}(Q_i))
$$
We need to define the projection of the groundstate
wavefunction
$$
\omega_1\otimes \omega_2 \in H^*(\CM^{\rm spl}(Q_1)\times \CM^{\rm spl}(Q_2) )
$$
onto a groundstate wavefunction in $H^*(\CM^{\rm spl}(Q_3)) $.

A conjecture, motivated by the work of Nakajima, and
of Ginzburg et. al., \nakajima\gkv,
is the following.
Suppose first that the three vectors $Q_i$
in \scati\
represent {\it positive} BPS states.
Recall that the charges are Chern characters of
sheaves. There is only one natural way that
the three sheaves $\CE_1, \CE_2, \CE_3$ can be related and
satisfy \scati. They must fit into an exact sequence:
\eqn\seqone{
0 \rightarrow \CE_1 \rightarrow
\CE_3 \rightarrow \CE_2 \rightarrow 0
}
or
\eqn\seqtwo{
0 \rightarrow \CE_2 \rightarrow
\CE_3 \rightarrow \CE_1 \rightarrow 0
}
The ambiguity between \seqone\ and \seqtwo\ is
resolved by the requirement that $\CE_3$
be semistable: since Chern characters are
additive the inequality \semistbl\
cannot hold for both $\CF=\CE_1$ and $\CF=\CE_2$.
\foot{We thank D. Morrison for a helpful remark
on this point.}

We define  the {\it correspondence region}
to be the subset of
$\CM(Q_1)\times \CM(Q_2)\times \CM(Q_3)$
defined by the set of triples:
\eqn\rrplus{
\eqalign{
\CC^{+++}(Q_1,Q_2;Q_3) & =
\{ (\CE_1, \CE_2, \CE_3) :
0 \rightarrow \CE_1 \rightarrow \CE_3 \rightarrow \CE_2 \rightarrow 0  \} \cr
}
}
If in \scat\scati\ we have some negative BPS
states then we rewrite \scati\ in terms of
positive vectors and write the corresponding
sequence. For example, suppose
$Q_1>0, Q_2<0, Q_3>0$. Then
$$
Q_1=Q_3 + (-Q_2)
$$
and the correspondence is
\eqn\rrmin{
\eqalign{
\CC^{+-+}(Q_1,Q_2;Q_3) &=
\{ (\CE_1, \CE_2, \CE_3) :
0 \rightarrow \CE_3 \rightarrow \CE_1 \rightarrow \CE_2 \rightarrow 0  \}  \cr
&
\subset  \CM (Q_1) \times \CM (-Q_2) \times \CM (Q_3) \cr}
}
when $\mu(\CE_3)\leq \mu(\CE_1)$, etc.
We can now state the

{\it Correspondence conjecture}:
 We conjecture that the residue of the boundstate pole
 is   the overlap of
the quantum wavefunctions on the correspondence
region:
\eqn\overlap{
\langle \omega_3 \vert \CR(\omega_1 \otimes \omega_2) \rangle
=
\int_{\CC(Q_1,Q_2;Q_3)} \omega_3^* \omega_1 \omega_2
}

{\bf Remarks}

\item{1.} In order for \overlap\ to make sense
we must first restrict the forms $\omega_i$ from
$\CM^{\rm spl}(Q_i)$ to the moduli space of
irreducible instantons $\CM^{\rm irred}(Q_i)$
(using the Donaldson-Uhlenbeck-Yau theorem).
We are then assuming that the forms extend
to the reducible locus, perhaps with singularities,
but such   that the integrals over
$\CM(Q)$ are well-defined.

\item{2.}
This definition of $\CR$ is extremely natural. It
amounts to the assumption that the bound state
formation simply corresponds to {\it local additivity of
the Chan-Paton vector spaces.}

\item{3.} The structure constants are therefore
given in terms of an intersection number, in
harmony with the topological field theory
interpretation.

\item{4.} Recall that the sheaf description
of 2-brane states required $[\Sigma]$ to
be of type $(1,1)$. Thus the above proposal
does {\it not} cover the scattering of 2-branes
which cannot be simultaneously made into
$(1,1)$ classes by rotation of complex structure.

\item{5.} The proposal  \overlap\ admits a nontrivial
consistency check in terms of the degree of the
form $\CR(\omega_1 \otimes \omega_2)$. See
appendix B for details.

\subsec{An heuristic argument for the correspondence
conjecture}

Here we try to justify further the correspondence
conjecture. The basic strategy is to use a standard
result of quantum mechanics: The residue of a
bound state pole is related to the coefficient of
the exponential falloff in the bound state
wavefunction
\ref\landau{See, e.g., L.D. Landau and
E.M. Lifschitz, {\it Quantum Mechanics},
Pergamon Press,  sec. 128.}.
We therefore attempt
 to construct an $L^2$ harmonic form
\eqn\scatstat{
\omega'_3 \in H^*(\CM'(Q_3),\IC)
}
with the asymptotic behavior:
\eqn\polform{
\omega'_3  \rightarrow e^{-\vert \vec p_{pole} \vert
\vert \vec a^{(1)} -  \vec a^{(2)}\vert } \omega_1
\omega_2
}
in the region in $\CM'(Q_3)$  corresponding
to widely separated states $\psi_1, \psi_2$
at positions $\vec a^{(i)}\in\IR^D$.
\foot{When the state is a  bound state at threshold
the falloff will be a power of $r=\vert
\vec a^{(1)} - \vec a^{(2)}\vert$.}
The restriction of $\omega_3'$ to
$\CM(Q_3)$ should define
 the product $\CR(\psi_1\otimes \psi_2)$.

Here we are assuming that we can restrict and
extend $H^*(\CM^{\rm spl}(Q)) \rightarrow
H^*(\CM^{\rm irred}(Q)) \rightarrow
H^*(\CM (Q))$ as in remark 1 above.
We are also assuming that there is a reasonable
Hodge theory on the
singular stratified space
$\CM'(Q)$ which allows us to
identify cohomology classes with harmonic
forms.

Now, to construct $\omega_3'$
we use the formulation of $\CM'(Q_3)$ in terms
of the generalized Hitchin system \ghs\
described above.
Recall that this space is stratified by the
partitions $S^r_\nu(\IR^D)$.

A. Suppose $Q_1=(r_1,\dots),
Q_2=(r_2,\dots)$. On all strata except
$\pi^{-1}(\Delta^{(r_1)} \times
\Delta^{(r_2)})$ and
$\pi^{-1} (\Delta^{(r_3)})$ we take
 $\omega_3' =0 $.

B. On the stratum $\pi^{-1}(\Delta^{(r_1)} \times
\Delta^{(r_2)})$ where we have the
Higgs field
\eqn\hggs{
\Phi_i = \pmatrix{a_i^{(1)} {\bf 1}_{r_1 \times r_1} & 0 \cr
0 & a_i^{(2)} {\bf 1}_{r_2 \times r_2} \cr}
}
we take
\eqn\strati{
\omega_3' = e^{- \vert \vec a^{(1)} - \vec a^{(2)} \vert \vert
\vec p_{pole} \vert } \omega_1 \omega_2
}
where $\omega_i$ are harmonic forms on
 $\CM^{\rm spl}(Q_i)$ restricted and extended to
$\CM(Q_i)$. Asymptotically the Hamiltonian
is written in terms of
the Laplacians on the moduli spaces
$\CM_1$, $\CM_2$ as:
$H= -{d^2 \over d\vec a^2} + \Delta_1 + \Delta_2$

C. Finally, on the stratum
$\vert \vec a^{(1)} - \vec a^{(2)} \vert =0$,
corresponding to $\pi^{-1} (\Delta^{(r)})$ we take:
\eqn\stratii{
\omega_3' =
\int_{\CM_1 \times \CM_2}
\eta(\CC \rightarrow \CM(Q_1) \times \CM(Q_2)
\times \CM(Q_3)) \omega_1  \omega_2
}
Here $\eta(X\rightarrow Y)$ denotes a harmonic
representative of the
Poincar\'e dual of a space $X$ embedded in
$Y$.

This defines a form $\omega_3'$ on all of
$\CM'(Q_3)$. It is a harmonic form on the various
smooth strata. It is also continuous because,
if $m_3$ is in the {\it reducible locus } of
$\CM(Q_3)$ then $\CR$ imposes
$\CE_3 = \CE_1 \oplus \CE_2$ so
$\omega_3' = \omega_1 \omega_2$, just
right to match to stratum B. We
presume that such a form is unique.

Thus, modulo the above caveats,
we conclude that $\omega_3'$ is the
form representing the bound state in the
scattering process. Now, if we want the
overlap with a bound state
$\omega_3 \in H^*(\CM^{irred}(Q_3))$
we recall that $\omega_3$ only has support in
stratum C. The overlap
$\langle \omega_3\vert \omega_3' \rangle$ is
exactly given by \overlap.

This concludes the argument for the
correspondence conjecture.

Remarks:

\item{1.} The above is an ansatz for the bound state
waveform. If there turns out to be a nontrivial
metric on $S^r(\IR^D)$ then the ansatz will have to
be modified.

\item{2.} We have not treated the singularities at the
reducible instantons with   care. This is related to
Assumption A of section 4.2. It is possible that the
moduli space is smoothed out at the reducible
locus due to short fundamental string degrees of
freedom, along the lines of \dkps\ and some
evidence for this is provided by
\dglsmoor. This could lead to  corrections to \overlap,
but these corrections should vanish
 in the large volume limit.

\subsec{Implications of Heterotic/IIA duality}

The above geometrical formulation of the
algebra of BPS states together with string
duality have important applications to
mathematics.

\subsubsec{Nakajima algebras}

Nakajima constructed algebras using correspondence
varieties exactly as in the correspondence conjecture.
While these definitions make sense for any algebraic
surface the resulting algebraic structures are
relatively unknown.

In particular, the answer for K3 was not hitherto
known. We see that type II/heterotic duality
together with the correspondence conjecture makes
an interesting prediction
\ref\crdts{The idea that Nakajima's construction
would lead to a GKM algebra with root lattice
$\Gamma^{20,4}$ first arose in a discussion
with R. Dijkgraaf, I. Grojnowski, and S. Shatashvili,
in Nov. 1994. }:

{\it Nakajima's construction
applied to the moduli space of   $U(r)$
``instantons''  for all $r\geq 0$
defines a generalized Kac-Moody
algebra whose root lattice is $\Gamma^{20,4}$
and which can be described explicitly as the
algebra of BPS states in the
heterotic string on $T^4$.
}

In the next two sections we will show how
two special cases of this statement
reproduce exactly Nakajima's results.

\subsubsec{Remark on a conjecture of
Gritsenko \& Nikulin }

In a fascinating paper, Gritsenko and
Nikulin \nikuliniii\ postulated the existence of
generalized Kac-Moody algebras whose simple
roots are associated to collections of elements
in the NEF cone of a K3 surface.  The algebra
of $(0,2)$-brane fiber bound states in a K3 fibration,
or of $(0,2,4)$ fiber bound states
provide examples of such GKM algebras, by virtue of
string/string duality. One of these algebras is
 probably the algebra needed for the
``Mirror conjecture'' stated in  \nikuliniii.
(It is possible that one will need to take a
quotient algebra of the algebra of BPS states.)
Indeed, the $(0,2,4)$ bound states, when
interpreted on the heterotic side,
have counting functions associated with
various threshold corrections, which involve
automorphic forms of the type entering in
Gritsenko and Nikulin's conjecture.

\newsec{Comparison of heterotic and type II algebras:
the Heisenberg algebras}

In the next two sections we compare the BPS algebras
of the type II and heterotic strings on $K3$ and $T4$ respectively.
We will restrict to the subspace of Narain moduli space where
we have the orthogonal decomposition:
\eqn\classgeom{
\Gamma^{20,4} = \Gamma^{1,1} \oplus
\Gamma^{19,3} \subset
\IR^{1,1}\oplus \IR^{19,3}
}
so we can describe the K3 in terms of classical geometry.
A waveform $\omega\in H^*(\CM^{\rm spl}(Q))$ for
$Q=(r, c_1, r-L)$ corresponds, on the heterotic side, to a
vertex operator with   matter of the   form:
\eqn\extra{
  e^{i E t(z,\bar z) }
\CP(\p^* x(z))
  e^{i  {1 \over  \sqrt{2} } ({(L-r)\over  V}  - r V )X(z)}
  \otimes
e^{i  {1 \over  \sqrt{2} } ( {(L-r)\over  V}  + r V)\tilde X(\bar z)}
e^{ i   c_1 \cdot  Y}(z,\bar z)
}
The notation is as follows. $V$ is the real parameter
of the lattice $\Gamma^{1,1}$. It corresponds to
a radius on the heterotic side, and to the volume of the
K3 on the IIA side. $x(z)$ stands for all 26
 holomorphic left-moving
coordinates. $(X,\tilde X)$ are left and right-moving
coordinates on $\IR^{1,1}$, while
$Y=(y,\tilde y)$ are left and right coordinates on
$\IR^{19,3}$.

\subsec{Type IIA description}

Nakajima's Heisenberg algebra
\nakajima
\ref\gron{I. Grojnowski, ``Instantons and
affine algebras I: the Hilbert scheme and
vertex operators,'' alg-geom/9506020.}
corresponds to the
scattering of $0$-branes off of a single $4$-brane
bound to collections of $0$-branes. Indeed, the
space of such BPS states is:
 \eqn\heisrep{
\eqalign{
\CH & \equiv
\oplus_{L\geq 0} \CH_{BPS}^{(1;0;1-L)}\cr
& \cong
\oplus_{L\geq 0} H^*(X_2^{[L]}) \cr}
}
where in the second line we have used
\ellzer.

Let
us consider a single four-brane bound to zerobranes
with total 0-brane charge $1-L_1$, and let us
scatter a 0-brane of charge $-L_2$.
The corresponding BPS
states are represented by homology classes:
\eqn\nkheisi{
[S_1] + [\tilde \Sigma_{L_2}]   \rightarrow [S_3]
}
where
\eqn\nakheis{
\eqalign{
[S_1] & \in H_*(\CM(1,0,1-L_1)) \cr
[\tilde \Sigma_{L_2}] & \in H_*(\Xi_{L_2}) \cr
[S_3] & \in H_*(\CM(1,0,1-L_3)) \cr}
}
and $L_3=L_1+L_2$.
Nakajima defines the operator
$\alpha^{\Sigma}_{-L_2}$ by
\eqn\nakheisi{
\eqalign{
\langle [S_3] \vert \alpha^{\Sigma}_{-L_2} \vert [S_1] \rangle
&  = \CC^{+++}\cap
\bigl[ S_1 \times \tilde \Sigma_{L_2} \times S_3 \bigr]  \cr
& = \int_{\CC(Q_1,Q_2;Q_3)} \eta_{S_1} \eta_{S_3}
\eta(\tilde \Sigma_{L_2} \rightarrow \Xi_{L_2}) \cr}
}
Recall that $\eta$ stands for the Poincar\'e dual.

In a similar way the absorption of a 0-brane
bound state of charge $+L_2$ is defined by
\eqn\nakheisii{
\eqalign{
\langle [S_3] \vert \alpha^{\Sigma}_{L_2} \vert [S_1] \rangle
&  = \CC^{+-+}\cap
\bigl[ S_1 \times \tilde \Sigma_{L_2} \times S_3 \bigr]  \cr
& = \int_{\CC^{+-+}(Q_1,Q_2;Q_3)} \eta_{S_1} \eta_{S_3}
\eta(\tilde \Sigma_{L_2} \rightarrow \Xi_{L_2}) \cr}
}

Nakajima has shown that the operators
$\alpha^I_n$ defined in this way as operators on
\heisrep\
 form a Heisenberg algebra
\eqn\nkheisiii{
[\alpha_n^I , \alpha_m^J] = c_n \eta^{IJ} \delta_{n+m,0}
}
where $\eta^{IJ}$ is the intersection pairing on
$H_*(X)$ and $c_n$ are constants.
Ellingsrud and Stromme \ellingsrud\
have been able to calculate $c_n$ using intersection
theory and find, remarkably,
\eqn\wow{
c_n = n (-1)^n
}
so the operators defined this way are
canonically normalized in the sense of
string theory!

\subsec{Heterotic description}

According to \extra\ the heterotic operators
$A_{L,\zeta}$
corresponding to zerobranes of charge $L$
have a left-moving matter piece given by
\eqn\leftmov{
A_{L, \zeta}^{\rm left} =
\zeta\cdot \p x e^{-i {L\over \sqrt{2} V} (t+X)}
}
where $(X,\tilde X)$ are left/right coordinates
on $\IR^{1,1}$, and $x$ runs over all 26 left-moving
coordinates. We have $\zeta\cdot k = 0 $ and
$\zeta\sim \zeta + \lambda k $ where $k$ is
the lightlike momentum in the exponential
in \leftmov. Thus, $\zeta$ span a 24-real
dimensional space.

The algebra of the operators $A_{n,\zeta}$ is
easily computed. The boost $\vec p=0$,
and we simply   find:
\eqn\heip{
 \CR(A_{n,\zeta} \otimes A_{n',\zeta'})
= n \zeta\cdot\zeta'  \delta_{n+n',0}{i   \over  \sqrt{2} V}
(\p t + \p X)
}
Note that it is important that
we are working within BRST cohomology.

The algebra \heip\ is definitely not a Heisenberg
algebra, but becomes closer to a Heisenberg
algebra when we consider the scattering of
$0$-branes off of
a 4-brane bound to zerobranes. Thus, we consider
the subspace of BPS states \heisrep\ from the
heterotic side.
When acting on the module $\CH$ we
find  that the operator
${i   \over  \sqrt{2} V} (\p t + \p X) $
acts as the {\it c-number} $-1$, so
that, acting on the module $\CH$
the operators $A_{n,\zeta}$ are
represented by:
 \eqn\lrgvol{
[A_{n,\zeta} , A_{n',\zeta'}] =  n' \delta_{n+n',0}
 \zeta\cdot\zeta'
 }
Thus we  recover Nakajima's Heisenberg algebra.

{\bf Remarks}.

\item{1.}There is a very close analogy of the
above  $0$-brane
operators with  DDF operators.

\item{2.} The above remarks can be generalized to
scattering off of wrapped 4-brane states with
charges $(r,c_1, r-L)$ at fixed $r,c_1$. In this
case the Heisenberg algebra becomes
$[A_{n,\zeta} , A_{n',\zeta'}] =  r n' \delta_{n+n',0}
 \zeta\cdot\zeta' $

\item{3.} The previous remark can be further
generalized: When the volume of K3 satisfies
$V=1$ then in fact the algebra of pure 0-brane
and 4-brane bound states is the algebra $w_{\infty}$
of area-preserving diffeomorphisms. Note
that at $V=1$ the $X$-CFT has the symmetry
of $\widehat{SU}(2)_1$ and we can define
primary fields
\eqn\winfti{
V_{J,m}(X) \equiv \biggl(\oint e^{-i \sqrt{2} X}\biggr)^{J-m}
e^{i \sqrt{2} J X}
}
of dimension $\Delta = J^2$ where $J=0,1/2, 1,\dots$ and
$m\leq \vert J\vert$. It follows immediately
from \refs{\kleb,\grndrng}
that the corresponding BPS multiplets $\Psi^{(+)}_{J,m}$
with  energy $E= \sqrt{2}\sqrt{J^2-1}$ satisfy:
\eqn\winftii{
\CR(\Psi^{(+)}_{J_1,m_1}
\otimes \Psi^{(+)}_{J_2,m_2}) = (J_2 m_1 - J_1 m_2)
\Psi^{(+)}_{J_1+J_2-1,m_1+m_2}
}

\subsec{Intuitive picture}

There is an extremely simple intuitive picture that
explains the noncommutativity of the algebra of
scattering 0-branes off 4-branes.  We now define
$\alpha^I_{L} $ for $L>0$ to be the operator on BPS
states obtained by absorbing a charge $+L$
$0$-brane and projecting onto the BPS state.
Similarly, $\alpha^I_{-L} $ is the operation of
absorbing a charge $-L$ $0$-brane and projecting
onto the BPS state.
The 4-brane wrapped around a K3 constitutes
the Heisenberg vacuum:
$\alpha_L \vert 0 \rangle =0$
because absorbing a charge $L$ $0$-brane
breaks supersymmetry. Moreover, $0$-branes
  of charge $L$ can only annihilate $0$-branes
of charge $-L$. Using these simple pictures one
can understand some aspects of the algebras.
Note that if we replace K3 by T4 then the algebra
is {\it not} highest weight since T4 does not
break any supersymmetry. This is in accord with
the fact that in the toroidally compactified
type II theory there are two BPS towers, one on the
left and one on the right.

\newsec{Comparison of heterotic and type II algebras:
the Affine Lie  algebras}

\subsec{Nakajima's construction}

In a famous set of papers Nakajima
\nakajima\
showed how to construct highest weight
representations of affine Lie algebras on
the cohomology spaces of quiver varieties
using the correspondences \rrplus\rrmin.
In particular, Nakajima associated
$\widehat {SU}(n)_r$ current algebra to the
moduli of $U(r)$ instantons on the
ALE space $X_n(\vec \zeta) $ which
is a resolution of $ \IC^2/\IZ_n$.
\foot{The $\vec \zeta$ are the hyperk\"ahler
moduli.}
In this section we will show how to recover
Nakajima's result from comparison of
BPS algebras via string/string duality
\foot{A suggestion that heterotic/IIA duality
might be the right arena to explain Nakajima's
affine Lie algebras associated to ALE manifolds
appeared in \vafa.
 A related suggestion had been
proposed independently by one of us
previously in unpublished work. }.

Let us consider a family
 $S(\epsilon)$ of
K3 surfaces degenerating to a surface
with an $ADE$ singularity.
Thus, in addition to \classgeom\ we
suppose that $\Gamma^{19,3}(\epsilon)$
degenerates to $\Gamma^{19,3}_*$ where
\eqn\esp{
(\Gamma(\lieg) ;\tilde 0_R) \subset \Gamma^{19,3}_*
}
for an ADE root lattice $\Gamma(\lieg)$.
For simplicity we will restrict our attention to
  $\lieg=A_{n-1}$.
 In this limit the two-spheres associated to
the roots of $\lieg$ shrink to zero size
so we simultaneously take a limit $V\rightarrow\infty$
of the $H^0\oplus H^4$ lattice $\Gamma^{1,1}(V)$
so that the area of the two-spheres $A=\epsilon^2 V$ is fixed.
The result is that as $\epsilon\rightarrow 0$
the K3 degenerates to
 an ALE space of type ADE.

As $\epsilon\rightarrow
0$ the
moduli space of instantons on the K3
  breaks up into   components
corresponding to the various components
of finite action instantons
on the ALE.  The latter have the topological
classification by $\rho$, a flat $U(r)$
connection on $S^3/\IZ_n$.
So we expect that
 as the K3 degenerates to an ALE with
finite area 2-spheres the cohomology of the
moduli space of instantons behaves as
\eqn\cpts{
H^*(\CM(v;S(\epsilon) ))
{}~{\buildrel \epsilon \rightarrow 0 \over  \rightarrow}~
\oplus_\rho H^*(\CM(v,\rho; X_n(\vec \zeta)))
}
The definition of correspondences
carries over to Nakajima's definition
so we should expect to recover Nakajima's
current algebras if we translate the above
degeneration to the heterotic side.
By \cpts,  a prediction of string/string
duality is that on the heterotic
side we should find all the highest weight
representations. We will verify this
below.  Nakajima associated highest weight
representations to the middle-dimensional
cohomology. We should take all the
cohomology. This is in keeping with the
heterotic description where it is evident that
there are many representations of $\hat\lieg_r$
in $\CH_{BPS}$.

\subsec{Recovering Affine Lie algebras}

Let us consider the heterotic algebra of
BPS states  under the degeneration
\eqn\degen{
\lim_{\epsilon\rightarrow 0}
\Gamma^{1,1}(A/\epsilon^2) \oplus
\Gamma^{19,3}(\epsilon)
}
described above.

We will consider algebras and modules
associated to
states with charge vectors
$Q=(r, c_1, r-L)$.  These have  matter
operators  given by \extra, as above.
In the limit $\epsilon \rightarrow 0$
we obtain BPS states with internal left-moving
vertex operator:
\eqn\liealg{
\eqalign{
J^{\vec \alpha} \leftrightarrow
& e^{ i \vec \alpha\cdot \vec y}  \epsilon_\alpha \cr
\vec H \leftrightarrow &-  i   \p \vec y  \cr}
}
for roots $\vec\alpha$ of $\lieg$.
\foot{We use a vector sign to denote a vector
in a Euclidean signature space.}
The algebra of these states
is simply the Lie algebra $\lieg$,
in the Cartan-Weyl basis. ($\epsilon_\alpha$ is a
cocycle factor.)

We now consider a larger algebra
obtained by adding generators with
left-moving matter:
\eqn\emin{
\eqalign{
J^{- \theta}_{ +1}
& \leftrightarrow
 e^{- i  {1 \over  \sqrt{2} } {1 \over  V} (t(z)+X(z))}
e^{i \vec \theta \cdot \vec y(z)} \cr
J^{\theta}_{  -1}
& \leftrightarrow
 e^{ + i  {1 \over  \sqrt{2} } {1 \over  V} (t(z)+X(z))}
e^{- i \vec \theta \cdot \vec y(z)} \cr
}}
where $\vec \theta$ is the highest root of
$\lieg$.
The product of the two states in \emin\ is
\eqn\extcomm{
\CR(J^\theta_{-1}\otimes J^{-\theta}_{+1}) \sim
-i  \vec \theta\cdot \p \vec y
+ {i \over  \sqrt{2}} {1\over  V}
(\p t + \p X)
}

While we do not obtain an affine Lie algebra in this
way we can introduce a   subspace of
BPS states  analogous
to \heisrep. We choose $r\in\IZ_+,  c_1\in \Gamma^{19,3}$
and define:
\eqn\affrep{
\CH_{r,  c_1} \equiv
 \sum_{\alpha\in (\Lambda_R(\lieg);0), L \geq
0}\CH_{BPS}^{(r;\alpha +   c_1; r-L)}
}
where we hold $r,c_1$ fixed.

 Now, when acting on the module
\affrep\
${i \over \sqrt{2} V} (\p t + \p X)$ is not a c-number,
but acting on the summand
$\CH_{BPS}^{(r;\alpha +   c_1; r-L)}$
it becomes multiplication by
\eqn\mult{
-\half (r-{L-r\over V^2})  - {1\over 2 V}
\sqrt{(rV + {L-r \over V})^2+ 2(\vec c_1^R)^2}
}
This is a complicated function, but, in the
$V\rightarrow \infty$ limit it becomes
simply $-r$, a c-number on the entire
module $\CH_{r,c_1}$. Thus,
in the $V\rightarrow \infty$ limit we have:
 \eqn\afflierel{
[J^\theta_{-1}, J^{-\theta}_{+1}] =
 \vec \theta \cdot \vec H -r
}
Similarly, we can easily compute:
\eqn\normgen{
\eqalign{
\bigl[
\bigl( i \vec \theta \cdot \p \vec y -
 {i \over  \sqrt{2}} {1\over  V}
(\p t + \p X) \bigr), J_{+1}^{-\vec \theta} \bigr]
& = 2 J_{+1}^{-\vec \theta}
\cr
\bigl[
\bigl( i \vec \theta \cdot \p \vec y -
 {i \over  \sqrt{2}} {1\over  V}
(\p t + \p X) \bigr), J_{-1}^{+\vec \theta} \bigr]
& = -2 J_{-1}^{+\vec \theta}\cr}
}
(this is valid for all volumes $V$).
Therefore,  the subalgebra of BPS states
generated by \emin\liealg\ acting on
the module \affrep\  is a
deformation of  the affine Kac-Moody
 algebra $\hat \lieg_r$
of level $r$.
In the $V \rightarrow \infty$ limit
the subalgebra generated by \emin\liealg\
becomes exactly $\hat \lieg_r$.
   \emin\liealg\  are the Serre generators
of the algebra.

\subsubsec{The representations}

We will now show that by choosing
$c_1$ appropriately we can obtain
all the integrable highest weight
representations in \affrep\ in
the large volume limit.  Since
 $J_{+1}^{-\vec \theta}$ lowers $L$
the modules \affrep\ are
always  highest weight representations.

States of type \extra\ come with degeneracy
  $p_{24}(N)$
where
\eqn\levnum{
N= r(L-r) + \half c_1^2 + 1 \qquad .
}
The highest weight state in a highest
weight representation should have
degeneracy one. This can be ensured
by choosing $L=r$ and $c_1^2=-2$.
 We do {\it not} assume that $c_1$ is purely
left-moving, although its projection to
the subspace $\Lambda_R(\lieg) \otimes \IR$
must be some weightvector $\vec \lambda$ since
$\Gamma^{20,4}$ is selfdual. At enhanced
symmetry points \esp\ we can obtain all
dominant weights by choosing suitable
$c_1$.
\foot{We have only verified this for $\lieg$ of
rank $\leq 3$, but believe it to be generally true.}
We choose
  a basis vector   $\Psi_{r,c_1,0}$ in this space.
Consider the state:
\eqn\nullvect{
(J_{-1}^{\vec \theta})^n \Psi_{r,c_1,0}
}
This state has charge vector $Q=(r,c_1+n \theta, n)$.
The smallest value of $n$ for which
$\half Q^2 + 1 <0$ (and hence, for which there
cannot be any BPS state) is:
\eqn\nullvcti{
n= r - \vec \lambda \cdot \vec \theta + 1
}
since $c_1\cdot\theta = -\vec \lambda\cdot\vec\theta$.
For this value of $n$ the vector
\nullvect\ must vanish. Thus, we interpret
\nullvect\ for $n$ given by
\nullvcti\ as the null vector of the
integrable highest weight representation
of level $r$ with weight $\vec \lambda$.
\foot{To complete the argument we should
show that the states do not vanish for smaller
values of $n$. We have not done so.}
Similarly, we may demand that the
action of $J_{+1}^{-\theta}$ produce
zero because we have violated the BPS
condition. This implies the inequality:
\eqn\inthigh{
\half (c_1 - \theta)^2 + r(r-1-r) + 1 < 0
}
or, equivalently:
\eqn\hwt{
\vec \lambda\cdot\vec \theta \leq r
}
which is just the integrable
highest weight condition
for the affine Lie algebra.

 In fact, the space \affrep\
forms a module of states under the sub-algebra
generated by BPS states \extra\
with charges $Q=(0;\vec \alpha; -N)$, for
$\vec \alpha\in \Lambda_R(\lieg)$.
These are $2$-branes bound to arbitrary
numbers of $0$-branes.
This is in principle a larger algebra than
the affine Lie algebra.

{\bf Remark}. A recent paper \gebnic\
appears to be closely related to the construction
of this section.

\newsec{Conclusions and future directions}

In this paper we have discussed the algebras of
BPS states associated to toroidally
compactified heterotic strings and to IIA strings
on K3 surfaces. However, as we have stressed,
the concept of a BPS algebra is quite general.
In particular, to {\it any} Calabi-Yau 3-fold
$X_3$ there are two
canonically associated algebras,
$\lieg^A(X_3), \lieg^B(X_3)$ defined by
the BPS algebras in the IIA and the IIB
theory. Moreover,  by quantum mirror
symmetry,
if $X_3, \tilde X_3$ are mirror pairs then
\eqn\mrror{
\lieg^A(X_3)\cong \lieg^B(\tilde X_3)\qquad .
}
Almost nothing is known
about these Calabi-Yau algebras.
The formulation in terms of correspondences
gives a {\it definition} of the $IIA$ algebra
but does not give an effective calculational
scheme. Moreover, there is at present no
analogous mathematical formulation
of the IIB algebra other than that provided
by \mrror.

We would like to stress that the CY algebras are
algebras of nonperturbative
dyonic BPS states in $d=4,\CN=2$
 type II compactifications.
Much remains to be  understood here.
First, in the $d=4,\CN=4$ theory we can
use 6-dimensional string/string duality
of heterotic/T6 with IIA/$K3\times T2$
to obtain the electric subalgebra. This
will be a GKM algebra with root lattice
$\Gamma^{22,6}$.
  Moreover, the results of \DVV\ suggest
that the full $d=4,\CN=4$ dyonic BPS algebra
should be a  GKM algebra.
\foot{We are using the term GKM algebra
loosely here. See note below.}

Moving on to more complicated
Calabi-Yau threefolds,
heterotic/IIA duality  for K3-fibered
Calabi-Yau's shows that the
$(0,2,4)$ fiber bound states  form
a GKM algebra. This suggests that the full
dyonic algebra
$\lieg^A(X_3)$ is   a generalized
Kac-Moody algebra and would fit in well
with  the natural conjecture that the result of
\DVV\ should generalize to $d=4,\CN=2$
compactifications.

There are many interesting avenues for
further investigation of the above
ideas.  It would be interesting to investigate
in detail the algebras of BPS states associated
to 4-folds of exceptional holonomy together
with their supersymmetric cycles.
In view of recent developments \toe\
the BPS algebra of
0-branes in IIA theory on tori seems
of particular importance.

In \hm\ it was shown that automorphic
forms of the kind appearing in the
study of GKM algebras appear in threshold
corrections. At the present moment
we do not have a good understanding
of how the algebra of BPS states and
algebras appearing in threshold
corrections are connected.
Indeed, until recently no clear
connection has been established between
any automorphic form appearing in
threshold corrections and a
concretely defined GKM.
That situation has been
improved recently \hmii,
but there is still much to learn.

Recently a very interesting phase
transition with an infinite number
of massless dyonic particles has
been discussed \refs{\vmii,\kmv}.
These states may be thought of as
analytic continuations of $(0,2,4)$
brane bound states in a Del Pezzo
surface. These will form a subalgebra of
the IIA algebra. Perhaps the phase
transition discussed in \vmii\kmv\ is
a phase - rather analogous to
$D=2$ string theory at the self-dual
radius, or to total string compactification
at special radii for the timelike coordinate,
  at which  an entire GKM gauge
algebra is becoming  unbroken.

\bigskip
\centerline{\it Notes added}

\item{1.} We would like to draw the reader's
attention to two  papers making use
of  correspondences in D-brane
interactions   \nakatsu\douglas.

\item{2.} R. Borcherds has pointed out to
us that our use of the term ``generalized Kac-Moody
algebras'' is inaccurate. The algebras
defined in \borchalg\borchi\ are $\IZ$ graded,
whereas our algebras are graded by a
lattice which is possibly non-Lorentzian
(e.g. $II^{p,q}$).  It seems to us to be an
important and interesting problem
 to develop the theory of
this more general class of algebras.

\bigskip
\centerline{\bf Acknowledgements}\nobreak

The remarks  on correspondences were
inspired by discussions and collaboration
with A. Losev,
N. Nekrasov, and S. Shatashvili on
 Nakajima algebras.
We would also like to thank T. Banks,
R. Dijkgraaf, M. Douglas, D. Freed.
M. Green, S. Katz, E. Martinec,
D. Morrison, R. Plesser,
J. Polchinski, G. Segal, A. Sen,
A. Strominger,   S. Shenker,
W. Taylor, A. Todorov, E. and H. Verlinde,
and G. Zuckerman
for discussions.

We thank the Rutgers physics department and
the ITP at Santa Barbara for hospitality during
the course of this work. GM would like to
thank  the Aspen Center for Physics for providing
a stimulating atmosphere during the completion
of this work. Some of these results were announced
on July 16, 1996 at the
conference Strings '96 at Santa Barbara. We
thank the organizers for the opportunity to
present them.
This work was  supported in part
by NSF Grant No.~PHY 91-23780 and
DOE grant DE-FG02-92ER40704.

\appendix{A}{The Hilbert scheme of points}

We collect here a few basic facts about the
Hilbert scheme of points. For more
information see \gottsche.

For any manifold $X$ we denote
\eqn\symmprod{
S^N(X) \equiv [X \times \cdots \times X]/S_N
}
This space has orbifold singularities whenever
two or more points in $X^N$ coincide. Indeed,
the space may be written as a stratified space
parametrized by the partitions of $N$:
\eqn\partenn{
S^N(X)  = \amalg_\nu (S^N(X) )_\nu
}
where if $\nu$ is the parition
$(1)^{n_1} (2)^{n_2} \cdots (s)^{n_s}$
then
$$
(S^N(X) )_\nu = \prod_i [X^{n_i} - BD]/S_{n_i}
$$
and $BD$ stands for the ``big diagonal'' where
any two points coincide.

If $X$ is a complex surface then there is a
resolution of singularities
$$
\pi: X^{[N]} \rightarrow S^N X
$$
given by the ``Hilbert scheme of points on $X$.''
This is the moduli space of sheaves supported
at points of length $N$. The length is given by
the dimension of the stalk at the point.

The Hilbert scheme of points on higher
dimensional complex manifolds  also
admits a projection to $S^NX$ but is
not smooth.

While the space $X^{[N]}$ parametrizes
sheaves supported at points it also parametrizes
sheaves of generic rank one which are rank zero  at a finite
set of points.
The parametrization is via
\eqn\shvv{
0 \rightarrow \CI \rightarrow \CO \rightarrow \CS \rightarrow 0
}
where $\CS$ is a skyscraper sheaf, and $\CI$ is the
sheaf defining the rank 1 Chan-Paton sheaf.

A local model for $\CI$ on the open
dense space $(X^N - BD)/S_N$
is the following.
Suppose we work near a point $x=y=0$. The
structure sheaf is just the sheaf of
analytic functions $\CO= \IC[[x,y]]$.
On the other hand,   $\CI(U)$ for $U$
containing  $x=y=0$ is the $\CO= \IC[[x,y]]$
module of analytic
series vanishing at $x=y=0$. Explicitly
we have:
\eqn\mxml{
\CI(U) = \{ a_{01} x + a_{10} y + a_{20} x^2 +
a_{11} xy + a_{02} y^2 + \cdots \}
}
on small open sets $U$ containing $x=y=0$.
The sequence \shvv\ gives the vector space
$\CS(U) = \{ a_{00} \} $ with $\CO$-module
structure $x\cdot a_{00}=y \cdot a_{00}=0$,
if $x=y=0$ is in $U$, that is, $\CS$ is a
skyscraper sheaf.  At the other
extreme, $\pi^{-1}(P) $ for $P\in \Delta^{(L)}$ (with
$\Delta^{(L)}$ the small diagonal)
parametrizes ideals $\CI$ in $\CO$ such that
$\CO/\CI$ is of dimensional $L$ and supported
at $P$.

\appendix{B}{Consistency check for the
correspondence conjecture}

It is possible to give a nontrivial consistency check
of \overlap\ by considering the degrees of the
forms involved. We thank G. Segal and K. Hori
for asking the
questions which led to this calculation.

A massive BPS multiplet transforms under
the group $SO(5)$,   the little group of
$\vec p=0$, as well as
the $d=6, \CN=2$ supertranslation
algebra. $SO(5)$ acts on the space of multiplets
and we expect the two Cartan generators to
be conserved in forming the BPS product.
Let us see how this is realized from the type IIA
side.

We consider the process \rrplus\ for definiteness,
and moreover suppose $Q_1\not= Q_2$. The
moduli space $\CM^{\rm spl}(Q)$ inherits a hyperk\"ahler
structure from $X_2$. As is well known,
the complex cohomology of a K\"ahler manifold
has an action of  $sl(2,\IC)$ \GrHa. On a
hyperk\"ahler  manifold this is promoted to $so(5,\IC)$
\ref\verbitsky{M. Verbitsky,
``Cohomology of compact
hyperk\"ahler manifolds and its applications,''
alg-geom/9511009}.
The two Cartan generators are diagonal
on $H^{p,q}(\CM(Q),\IC)$ and take the values:
\eqn\jays{
\eqalign{
J_{12}  & = p+q - \dim_\IC \CM(Q) \cr
J_{34}   & = p-q  \cr}
}
It is natural to interpret the $so(5,\IC)$ action
as the action of the complexified little group,
see, e.g.,
\ref\emmeff{E. Witten, ``Phase transitions in
$M$-theory and $F$-theory,'' hep-th/9603150}.
Let us verify that these are conserved by
the product \overlap.

First, since $\CC$ is an analytic subvariety
its Poincar\'e dual is of type $(p,p)$. Hence
$J_{34}$ is trivially conserved.  It takes more
work to verify that $J_{12}$ is conserved.
To begin, the conservation of $J_{12}$ is
equivalent to:
\eqn\lefshetz{
\deg \omega_3 = \deg \omega_1 + \deg \omega_2
+ 2 Q_1 \cdot Q_2 -2
}
Now, \overlap\ predicts that
\eqn\cordeg{
\deg \omega_3 = \deg \omega_1 + \deg \omega_2
+ \dim_{\IR} \CM(Q_3) - \dim_\IR \CC^{+++}
}
since $\eta$ is a Poincar\'e dual. Now,
$ \dim \CC^{+++}$ can be computed as
\eqn\dmcurlyc{
\dim_\IC \CC^{+++} =
\dim_\IC \CM(Q_1) + \dim_\IC \CM(Q_2)
+ \dim_\IC  H^1(X; Hom(\CE_2, \CE_1)) -1
}
The reason for this is that, having chosen
$\CE_1, \CE_2$ there is a nontrivial space of
extensions given by $H^1(X; Hom(\CE_2, \CE_1))$.
(See, for example, \DoKro, Prop. 10.2.4.) The
extra $-1$ comes about because
$H^1(X; Hom(\CE_i, \CE_i))$ for $i=1,2$
are both 1-dimensional (since the sheaves
are simple) so the ``ratio'' of these
defines a spurious direction in
$H^1(X; Hom(\CE_2, \CE_1))$.
In terms of transition functions:
\eqn\trsnfun{
\pmatrix{\lambda_1 & 0 \cr
0 & \lambda_2 \cr} \pmatrix{1 & \chi_{\alpha \beta}\cr
0 & 1 \cr} \pmatrix{\lambda_1 & 0 \cr
0 & \lambda_2 \cr}^{-1}
=
\pmatrix{1 &
\lambda_1 \lambda_2^{-1} \chi_{\alpha \beta}\cr
0 & 1 \cr}
}
identifies a representative $\chi_{\alpha \beta}$ of
$H^1(X; Hom(\CE_2, \CE_1))$ with
$\lambda_1 \lambda_2^{-1} \chi_{\alpha \beta}$.
Thus, the RHS of \cordeg\ becomes
\eqn\cordegi{
  \deg \omega_1 + \deg \omega_2 + 4 Q_1 \cdot Q_2
- 2 \dim_\IC H^1(X; Hom(\CE_2, \CE_1)) -2
}
Now, by RRG we can compute
\eqn\dmach{
\eqalign{
 \dim_\IC H^1(X; Hom(\CE_2, \CE_1))
& = Q_1 \cdot Q_2 + \dim_\IC H^0(X; Hom(\CE_2, \CE_1)) \cr
& +
\dim_\IC H^2(X; Hom(\CE_2, \CE_1))\cr}
}
See, \mukvb, eq. 3.21. Moreover,
$H^2(X; Hom(\CE_2, \CE_1))$ should vanish
if $\CC$ is smooth (this is the space of obstructions).
Finally, from the long exact sequence associated to
\eqn\ses{
0 \rightarrow Hom(\CE_2, \CE_1)
\rightarrow Hom(\CE_2, \CE_3)
\rightarrow Hom(\CE_2, \CE_2)
\rightarrow 0
}
we get
$H^0(X; Hom(\CE_2, \CE_1))
\cong H^0(X; Hom(\CE_2, \CE_3))$.
However, since $\CE_2, \CE_3$ are
semistable one finds the latter space is
the zero vector space. Proof: If
$\psi: \CE_2 \rightarrow \CE_3$ were nonzero
then, since $\CE_2$ is semistable,
$\mu(\CE_2) < \mu(\CE_2/\ker\psi)$.
We also know $\mu(\CE_3) < \mu(\CE_2)$
but $\CE_2/\ker \psi \cong im(\psi) \subset \CE_3$
is a subsheaf of $\CE_3$, but this contradicts
semistability of $\CE_3$.  Thus
$ \dim_\IC H^1(X; Hom(\CE_2, \CE_1))
= Q_1 \cdot Q_2$
and we finally get agreement between the
RHS of \cordeg\ and \lefshetz.

\listrefs
\bye